\documentclass[aps,prd,superscriptaddress,nofootinbib,preprintnumbers,floatfix,tightenlines,fleqn,notitlepage,10pt]{revtex4-1}
\usepackage{amsmath,amsfonts,amssymb,amscd,amsxtra,amsthm}
\usepackage{graphicx}  
\usepackage{epstopdf}
\usepackage{dcolumn}  
\usepackage{bm}          
\usepackage{slashed}
\usepackage{cancel}
\usepackage{mathtools} 
\usepackage{amsbsy}
\usepackage{amstext}

\usepackage[utf8]{inputenc}
\usepackage{booktabs}
\usepackage[normalem]{ulem}
\usepackage[dvipsnames]{xcolor}
\usepackage{tabularx}
\usepackage{enumitem}
\usepackage{array}
\usepackage{tikz}
\usepackage{pgfplots}
\pgfplotsset{compat=1.18}  
\usepackage{multirow}
\usepackage{physics}
\usepackage{hyperref}
\usepackage{cleveref}

\renewcommand\sout{\bgroup \color{red} \ULdepth=-.5ex \ULset}

\begin{document}
\preprint{INHA-NTG-12/2026}
\title{A doubly charmed strange molecule from 
  coupled-channel dynamics} 

\author{Hee-Jin Kim}
\email[ E-mail: ]{hjin.ghim@gmail.com}
\affiliation{Quantum AI Innovative Technology Institute (Q-AIT),
  Hoseo University, Asan 31499, Republic of Korea} 
\author{Hyun-Chul Kim}
\email[ E-mail: ]{hchkim@inha.ac.kr}
\affiliation{Department of Physics and Institute of Quantum Science,
  Inha University, Incheon 22212, Republic of Korea} 
\date{\today}

\begin{abstract}
We investigate a doubly charmed molecular state $T_{cc\bar{s}}$ with
$C=+2$, $\mathcal{S}=+1$, and $I=1/2$ in an off-shell coupled-channel formalism.
Constructing the interaction kernel from a meson-exchange model
based on an effective Lagrangian and solving the coupled-channel
scattering equation in the Blankenbecler-Sugar reduction,
we find a pole only for $J^P=1^+$ among $J^P=0^\pm$, $1^\pm$, and
$2^\pm$. With the reduced cutoff mass of $600$~MeV, we obtain a
mass of $3972.90$~MeV, which corresponds to a binding energy of
$4.01$~MeV relative to the $D^*D_s$ threshold. Examining the relative
signs and magnitudes of the $S$-wave channel couplings extracted from
the pole residues, we demonstrate that the pole belongs predominantly
to the triplet ($\bm{3}_A$) representation. We also discuss the relevance of heavy-quark spin
symmetry to the bound state.    
\end{abstract}
\maketitle
\section{Introduction}
The LHCb Collaboration announced the first observation of the doubly
charmed tetraquark $T_{cc}(3875)^+$ in the $D^0D^0\pi^+$ mass
spectrum~\cite{LHCb:2021vvq, LHCb:2021auc}. Its minimal quark content
is $cc\bar{u}\bar{d}$, which shows that the state is manifestly
exotic, whereas some neutral hidden-charm states share their quantum
numbers with charmonia (see the reviews~\cite{Lebed:2016hpi,
  Guo:2017jvc, Brambilla:2019esw, Chen:2022asf, Meng:2022ozq}). The
state lies just below the $D^{*+}D^0$ threshold. A Breit-Wigner (BW)
fit gives a mass difference of $\delta m_{\mathrm{BW}}=-273$~keV
relative to this threshold and a width of
$\Gamma_{\mathrm{BW}}=410$~keV. A unitarized analysis yields
$\delta m_{\mathrm{pole}}=-360$~keV and
$\Gamma_{\mathrm{pole}}=48$~keV~\cite{LHCb:2021auc}, with the favored
quantum numbers $I(J^P)=0(1^+)$. The small binding energy and the
narrow width suggest a loosely bound $DD^*$ molecule.
Since the LHCb finding, a great deal of work has been devoted to the 
structure and production mechanism of the $T_{cc}^+$. These studies
focus mainly on the interaction that binds $DD^*$ and on the
observables that distinguish a molecule from a compact state. The
line shapes and decay properties of $T_{cc}$ have been analyzed in
Refs.~\cite{Albaladejo:2021vln, Du:2021zzh, Dai:2021wxi,
  Feijoo:2021ppq, Fleming:2021wmk, Ling:2021bir}. The binding has
been studied in meson-exchange models~\cite{Ren:2021dsi, Ke:2021rxd,
  Zhao:2021cvg, Asanuma:2023atv, Kim:2025ado}, in a molecular
description based on quark-level interactions~\cite{Chen:2021cfl},
and in a constituent quark model that includes one-pion
exchange~\cite{Meng:2023for}. Lattice QCD calculations of $DD^*$
scattering~\cite{Padmanath:2022cvl, Lyu:2023xro, Meng:2024kkp,
  Whyte:2024ihh} have examined the interaction independently of
experiment, and the dependence of the pole on the light-quark
mass~\cite{Abolnikov:2024key} connects these calculations to the
physical point. Femtoscopic correlation
functions~\cite{Kamiya:2022thy, Vidana:2023olz} provide a further
measurable test of these exotic states.

The molecular picture has also been applied to hidden-charm states
that contain a strange quark. The BESIII Collaboration observed the
$Z_{cs}(3985)^-$ in the $K^+$ recoil-mass spectrum of
$e^+e^-\to K^+(D_s^-D^{*0}+D_s^{*-}D^0)$~\cite{BESIII:2020qkh}, and
the LHCb Collaboration reported the $Z_{cs}(4000)^+$ and
$Z_{cs}(4220)^+$ in the $J/\psi K^+$ decay
mode~\cite{LHCb:2021uow}. The $Z_{cs}(3985)$ and $Z_{cs}(4000)$ have
been interpreted as $D_s^{(*)}\bar{D}^{(*)}$ molecules in
meson-exchange models~\cite{Yan:2021tcp}, in coupled-channel analyses
with contact interactions~\cite{Baru:2021ddn, Du:2022jjv}, in
quark-exchange calculations of their decays~\cite{Han:2022fup}, and
in a chiral potential approach at next-to-leading
order~\cite{Zhai:2022ied}. However, they have also been explained as
compact tetraquarks~\cite{Wang:2020iqt, Maiani:2021tri} or as
threshold cusps~\cite{Nakamura:2023swt}. Whether the $Z_{cs}(3985)$
and $Z_{cs}(4000)$ are the same state also remains
unresolved~\cite{Meng:2021rdg}. Strange hidden-charm pentaquarks show
the same pattern. The $P_{c\bar{c}s}(4459)^0$~\cite{LHCb:2020jpq} and
$P_{c\bar{c}s}(4338)^0$~\cite{LHCb:2022ogu} were found in the
$J/\psi\Lambda$ channel near the $\Xi_c\bar{D}^*$ and $\Xi_c\bar{D}$
thresholds, respectively. Evidence for the former has since been
reported by the Belle Collaboration~\cite{Belle:2025pey}. Both states
emerge as $\Xi_c^{(\prime,*)}\bar{D}^{(*)}$ molecules in
coupled-channel calculations~\cite{Wang:2022mxy, Feijoo:2022rxf,
  Clymton:2025hez}. If the same dynamics binds these meson-baryon
systems, we expect it also to bind two charmed mesons into a strange
counterpart of the $T_{cc}$  within the same picture.

Replacing $\bar d$ by $\bar s$ in $cc\bar u\bar d$ gives the
configuration $cc\bar u\bar s$ with $C=+2$ and $\mathcal{S}=+1$, which is
denoted by $T_{cc\bar s}$. The relevant two-body channels include
$DD_s$, $D^*D_s$, $DD_s^*$, and $D^*D_s^*$. For $J^P=1^+$, the two
lowest channels, $D^*D_s$ and $DD_s^*$, lie only $2.5$~MeV apart and
couple in $S$ wave. No channel opens below them, since parity
excludes $DD_s$. A pole a few MeV below these thresholds would then
be a shallow bound state with no two-body decay channel. The two
channels are distinct even in the isospin-symmetric limit, since the
vector meson is the nonstrange $D^*$ in one channel and the strange
$D_s^*$ in the other. The two thresholds are nearly degenerate because
the hyperfine splittings between $D^*$ and $D$ and between $D_s^*$ and
$D_s$ are similar. The $D^*D_s^*$ threshold lies about $144$~MeV
higher, but this channel cannot be neglected. In the limit of infinite
heavy-quark mass, the pseudoscalar and vector mesons with the same
light flavor form a spin doublet~\cite{Isgur:1989vq, Neubert:1993mb},
so heavy-quark spin symmetry (HQSS) relates the $D^*D_s^*$ interaction 
to those in the $D^*D_s$ and $DD_s^*$ channels. A coupled-channel
calculation is therefore required to quantify the contribution of
$D^*D_s^*$ to the binding. Moreover, since $D$
($D^*$) and $D_s$ ($D_s^*$) are distinguishable, Bose-Einstein
symmetry does not apply.

The interaction also differs from that in the nonstrange $T_{cc}$.
One-pion exchange, which provides the long-range attraction in
$DD^*$, is forbidden between $D^{(*)}$ and $D_s^{(*)}$. Any binding
must therefore come from the shorter-range exchanges of $\eta$, $K$,
$K^*$, $\sigma$, and $f_0$. In the present work, we thus investigate
the $T_{cc\bar{s}}$ system with $J^P=0^\pm$, $1^\pm$, and $2^\pm$.
The kernels are built from these shorter-range meson exchanges, with
the couplings taken from an effective Lagrangian that respects HQSS, chiral symmetry, and hidden local
symmetry. Inserting these kernels into the coupled Blankenbecler-Sugar
(BbS) scattering equations~\cite{Blankenbecler:1965gx,Aaron:1968aoz},
we compute the
off-shell transition amplitudes relevant to the $T_{cc\bar{s}}$. For
each $J^P$, we include the allowed transitions among $DD_s$,
$D^*D_s$, $DD_s^*$, and $D^*D_s^*$ and derive the partial-wave
amplitudes from $t$- and $u$-channel exchanges. The present off-shell
coupled-channel framework has already been applied successfully to
various resonances, such as light axial-vector
mesons~\cite{Clymton:2022jmv,Clymton:2023txd,Clymton:2024pql},
heavy-meson molecules~\cite{Kim:2023htt,Kim:2025ado}, and hidden-charm
pentaquarks~\cite{Clymton:2024fbf,Clymton:2025hez,Clymton:2025zer}.

\begin{figure*}[t]
  \centering
  \includegraphics[width=0.95\textwidth]{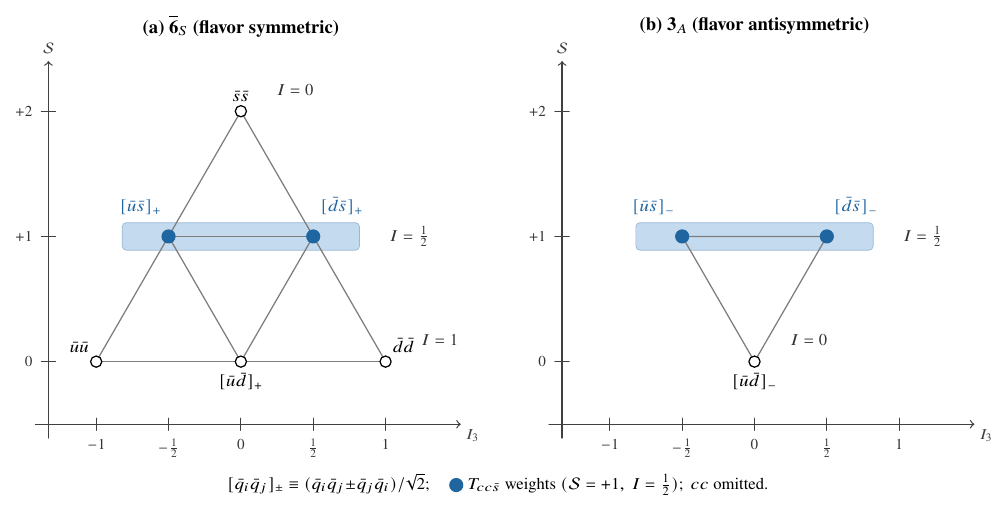}
  \caption{Weight diagrams of the symmetric $\bar{\bm{6}}_S$ and
  antisymmetric $\bm{3}_A$ representations in
  $\bar{\bm{3}}\otimes\bar{\bm{3}}=\bar{\bm{6}}_S\oplus\bm{3}_A$. The
  $cc$ pair is a light-flavor singlet and is therefore omitted. The
  horizontal and vertical axes denote $I_3$ and strangeness $\mathcal{S}$,
  respectively. Filled symbols highlight the $\mathcal{S}=+1$, $I=1/2$ weights:
  both representations contain a $T_{cc\bar{s}}$ isodoublet, although
  their light-antiquark wave functions have opposite exchange
  symmetry.}
  \label{fig:1}
\end{figure*}

\begin{figure}[htbp]
  \centering
  \includegraphics[scale=0.95]{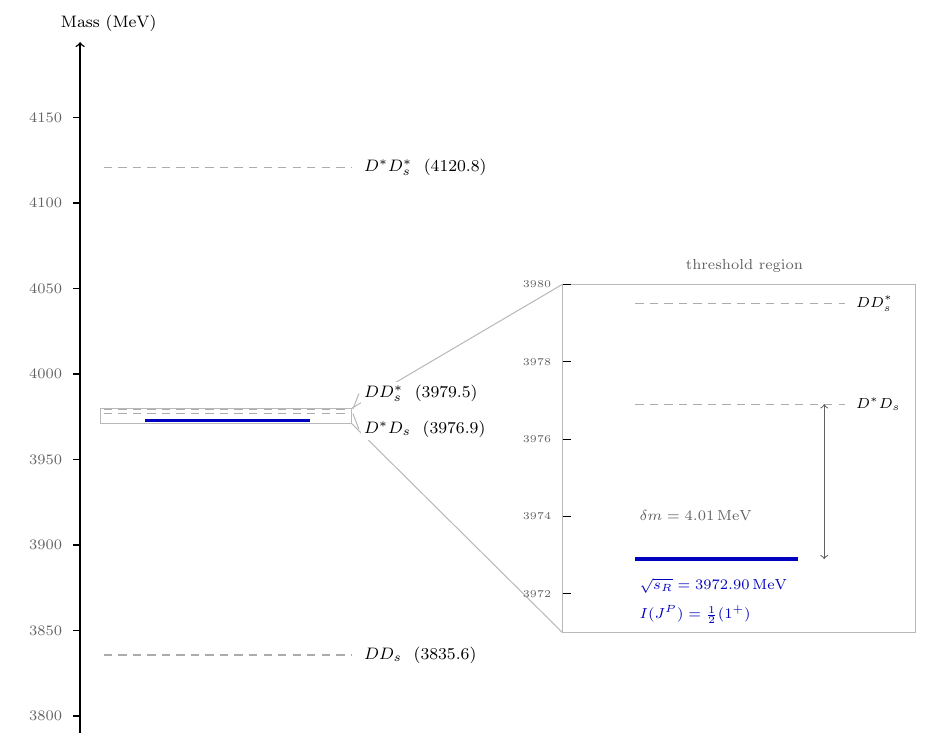}
  \caption{Open-charm channel thresholds of the $T_{cc\bar{s}}$ system
  (dashed lines, in MeV) and the bound-state pole obtained at
  $\Lambda_0=600$~MeV (solid line). The inset magnifies the threshold
  region, resolving the $2.5$~MeV splitting between the $D^*D_s$ and
  $DD_s^*$ thresholds and the binding energy of $4.01$~MeV relative
  to the former.}
  \label{fig:2}
\end{figure}

From the transition amplitudes obtained in the present work, we
examine two main points: which channels are required for binding, and
to which flavor representation the pole belongs. For $J^P=1^+$, we
compare the full three-channel solution with each of its two-channel
subsystems. For the flavor content, we use the SU(3) decomposition of
the two light antiquarks,
$\bar{\bm{3}}\otimes\bar{\bm{3}}=\bar{\bm{6}}_S\oplus \bm{3}_A$. The
nonstrange charmed mesons form isodoublets, and their strange
counterparts are isosinglets, so all four channels have $I=1/2$. Both
the symmetric $\bar{\bm{6}}_S$ and the antisymmetric $\bm{3}_A$
representations contain a multiplet with $\mathcal{S}=+1$ and $I=1/2$, as shown
in Fig.~\ref{fig:1}. Isospin therefore does not distinguish these two
flavor configurations, and symmetry alone fixes neither the number of
poles nor their assignment to a representation, but the
coupled-channel dynamics does. We identify the antisymmetric
$\bm{3}_A$ as the dominant representation from the relative signs and
magnitudes of the $S$-wave couplings, which we extract from the
diagonal and off-diagonal residues at the pole. This result indicates
that the $T_{cc\bar{s}}$ found in the present work may belong to the
same SU(3) multiplet as the isoscalar $T_{cc}$.
Using the reduced cutoff mass $\Lambda_0=600$~MeV, we find a pole
only for $J^P=1^+$, about $4$~MeV below the $D^*D_s$ threshold, as
summarized in Fig.~\ref{fig:2}. We obtain no such pole from any of
the two-channel subsystems with the same interaction parameters, so
binding requires all three channels, $D^*D_s$, $DD_s^*$, and
$D^*D_s^*$.

This paper is organized as follows. Section~\ref{sec:2} describes
the BbS equation and the interaction kernel. Section~\ref{sec:3}
presents the channel-coupling comparisons and pole properties,
positive-parity cross sections, cutoff dependence, and flavor
classification. Section~\ref{sec:4} summarizes the results.

\section{Theoretical Framework}
\label{sec:2}
\subsection{Coupled-channel equation with three-dimensional
reduction} 
\label{sec:2-1} 
The $S$-matrix elements are related to the
transition amplitude $\mathcal{T}_{fi}$ through the conservation of
four-momentum:
\begin{align}
S_{fi} = \delta_{fi} - i(2\pi)^4 \delta^{(4)}(P_f-P_i) \mathcal{T}_{fi},
\label{eq:1}
\end{align} 
where $P_{i(f)}$ denotes the total four-momentum of the initial
(final) hadronic state. This formulation isolates the nontrivial
scattering dynamics into the amplitude $\mathcal{T}_{fi}$. We use
relativistically normalized one-particle states,
$\langle\bm p'|\bm p\rangle=2E_{\bm p}(2\pi)^3
\delta^{(3)}(\bm p'-\bm p)$. The overall sign in
Eq.~\eqref{eq:1} fixes our amplitude convention. When the system is
restricted to two-hadron states, the transition amplitude is obtained
by solving the coupled-channel Bethe-Salpeter (BS) equation
illustrated in Fig.~\ref{fig:3}:
\begin{figure}[htbp]
  \centering
  \includegraphics[width=0.76\textwidth]{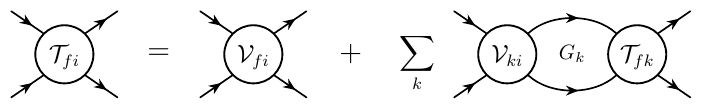}
  \caption{Graphical representation of the coupled-channel
  Bethe-Salpeter equation~\eqref{eq:2}. The sum runs over intermediate
  two-body channels $k$, with loop integration implicit.}
  \label{fig:3}
\end{figure}
\begin{align}
\mathcal{T}_{fi}(p',p;s) &= \mathcal{V}_{fi}(p',p;s) + \frac{1}{(2\pi)^4}
  \sum_k \int d^4q_k\, \mathcal{V}_{fk}(p',q_k;s) 
  G_k(q_k;s) \mathcal{T}_{ki}(q_k,p;s),
  \label{eq:2} 
\end{align}
where $\sqrt{s}$ is the total energy in the center-of-mass (CM) frame.
The variables $p$ and $p'$ represent the relative four-momenta of the
initial state $i$ and final state $f$, respectively. The index $k$
runs over intermediate two-body channels allowed by the relevant
symmetries, and $q_k$ is the off-shell relative four-momentum of the
intermediate state $k$.

In the present work, the kernel $\mathcal{V}_{fi}$ of the integral
equation~\eqref{eq:2} is constructed from the Feynman amplitudes for
the allowed two-body transitions, which are discussed in detail in
Sec.~\ref{sec:2-2}. The two-body factor $G_k$ is given by
\begin{align}
i G_k = \,\frac{i}{(q_1^k)^2-(m_1^k)^2}
             \frac{i}{(q_2^k)^2-(m_2^k)^2},
\label{eq:3}
\end{align}
where $m_1^k$ and $m_2^k$ are the masses of the hadrons in channel
$k$. The meson masses used in this work are listed in
Table~\ref{tab:1}. We adopt isospin-averaged masses for $D$ and $D^*$
based on the PDG listings~\cite{PDG:2024cfk}.

\setlength{\tabcolsep}{10pt}
\renewcommand{\arraystretch}{1.5}
\begin{table}[htp]
\centering
\caption{Particle masses used in the present work.}
\begin{tabular}{c|c}
\hline\hline
Particle & Mass [MeV] \\
\hline
$D$ & 1867.25 \\
$D^*$ & 2008.56 \\
$D_s$ & 1968.35 \\
$D_s^*$ & 2112.20 \\
$\eta$ & 547.862 \\
$K$ & 495.0 \\
$K^*$ & 893.9 \\
$\sigma$ & 500.0 \\
$f_0(980)$ & 990.0 \\
\hline\hline
\end{tabular}
\label{tab:1}
\end{table}

A three-dimensional reduction is usually employed to solve the BS
equation numerically. A standard choice is the BbS
scheme~\cite{Blankenbecler:1965gx, Aaron:1968aoz}, in which the two-body factor of Eq.~\eqref{eq:3} is
replaced by a constrained propagator that fixes the relative energy of
the intermediate state:
\begin{align}
G_k(q) = \frac{\pi}{\omega_1^k \omega_2^k} \delta\left(q^0 - 
\frac{\omega_1^k-\omega_2^k}{2}\right) 
\frac{\omega_1^k+\omega_2^k}{s-(\omega_1^k+\omega_2^k)^2+i\epsilon}, 
\label{eq:4}
\end{align}
where $\omega_{1(2)}^k=\sqrt{\bm{q}^2+(m_{1(2)}^k)^2}$ denote the
on-shell energies of the two particles in channel $k$. This reduction leads
to the three-dimensional BbS equation:
\begin{align}
T_{fi}(\bm{p'},\bm{p};s)=V_{fi}(\bm{p'},\bm{p};s)+ \sum_k \int d^3q \, 
  V_{fk}(\bm{p'},\bm{q};s) \tilde{G}_k(\bm{q}) T_{ki}(\bm{q}, \bm{p};s)
\label{eq:5}  
\end{align}
with the reduced propagator
\begin{align}
\tilde{G}_k(\bm{q}) = \frac{1}{(2\pi)^3} 
\frac{\omega_1^k+\omega_2^k}{2\omega_1^k\omega_2^k} 
\frac{1}{s-(\omega_1^k+\omega_2^k)^2+i\epsilon}.
\label{eq:6}  
\end{align}
Within the BbS reduction, the off-shell dynamics is retained through
the dependence of $V$ and $T$ on the intermediate relative momentum
$\bm q$, which is integrated over in Eq.~\eqref{eq:5}. The
integration samples momenta away from the two-body energy shell
$\sqrt{s}=\omega_1^k+\omega_2^k$.

We start from the helicity amplitudes, where each hadron is assigned a
specific helicity $\lambda_i$. The partial-wave decomposition is
performed using Wigner $d$-functions, $d_{\lambda\lambda'}^J(\theta)$,
where $J$ is the total angular momentum, and $\lambda$ ($\lambda'$)
represents the helicity difference between the particles in the
initial (final) state. Amplitude indices
$\boldsymbol\lambda=(\lambda_1,\lambda_2)$ and $\boldsymbol\lambda'$
indicate the individual particle helicities. The integral equation is
projected onto the partial-wave matrix elements as follows:
\begin{align}
T_{\boldsymbol\lambda'\boldsymbol\lambda}^{J(fi)} (\mathrm{p}',\mathrm{p};s) &=
  V_{\boldsymbol\lambda'\boldsymbol\lambda}^{J(fi)}(\mathrm{p}',\mathrm{p};s) +
  \sum_{k,\boldsymbol\lambda_k} \frac{1}{(2\pi)^3} \int_0^\infty d\mathrm{q}_k \,
  \frac{\omega_1^k+\omega_2^k}{2\omega_1^k\omega_2^k}
  \frac{\mathrm{q}_k^2 V_{\boldsymbol\lambda'\boldsymbol\lambda_k}^{J(fk)}
  (\mathrm{p}',\mathrm{q}_k;s) T_{\boldsymbol\lambda_k\boldsymbol\lambda}^{J(ki)} 
  (\mathrm{q}_k,\mathrm{p};s)}{s-(\omega_1^k+\omega_2^k)^2+i\epsilon}, 
\label{eq:7}
\end{align}
where the partial-wave kernel $V^J$ and transition amplitude $T^J$ are
defined by the following projections:
\begin{align}
V_{\boldsymbol\lambda'\boldsymbol\lambda}^{J(fi)}(\mathrm{p}',\mathrm{p};s) &= 2\pi \int_{-1}^1
  d\cos\theta \, V_{fi}(\mathrm{p}',\mathrm{p},\theta;s)  
  d_{\lambda\lambda'}^J(\theta),  \nonumber \\
T_{\boldsymbol\lambda'\boldsymbol\lambda}^{J(fi)} (\mathrm{p}',\mathrm{p};s) &= 2\pi \int_{-1}^1
  d\cos\theta \, T_{fi}(\mathrm{p}',\mathrm{p}, \theta;s) 
  d_{\lambda\lambda'}^J(\theta).
\label{eq:8}  
\end{align}
$\theta$ denotes the scattering angle between the three-momenta
$\bm{p}$ and $\bm{p}'$, while $\mathrm{p}$, $\mathrm{p}'$, and
$\mathrm{q}_k$ stand for the magnitudes of the respective
three-momenta. The first momentum argument represents the final
momentum $\mathrm{p}'$ and the second the initial momentum
$\mathrm{p}$. The symbols $T,V$ and $\mathcal T,\mathcal V$ 
refer to the same amplitude normalization. No momentum or flux factors
are absorbed when passing to the three-dimensional or partial-wave
notation. 

For numerical treatment, it is preferable to isolate the pole in the
two-body propagator. The singularity occurs at the on-shell momentum
$\tilde{\mathrm{q}}_k$, where $\sqrt{s} = \omega_1^k + \omega_2^k$. By
defining $\mathcal{F}(\mathrm{q})$ such that the ratio
$[\mathcal{F}(\mathrm{q}) - \mathcal{F}(\tilde{\mathrm{q}}_k)] /
[\sqrt{s} - (\omega_1^k + \omega_2^k)]$ remains finite, we can
separate the singular part from the regular part of the integral. The
singular part is evaluated as a principal-value integral, which gives
\begin{align}
  T^{J(fi)}_{\boldsymbol\lambda'\boldsymbol\lambda} (\mathrm{p}',\mathrm{p}) = 
  V^{J(fi)}_{
  \boldsymbol\lambda'\boldsymbol\lambda} (\mathrm{p}',\mathrm{p}) + \frac{1}{(2\pi)^3}
  \sum_{k,\boldsymbol\lambda_k}\left[\int_0^{\infty}d\mathrm{q}
  \frac{\mathrm{q}\omega_k}{\omega_1^k\omega_2^k}\frac{\mathcal{F}
  (\mathrm{q})-\mathcal{F}(\tilde{\mathrm{q}}_k)}{s-\omega_k^2}+ 
  \frac{1}{2\sqrt{s}}\left(\ln\left|\frac{\sqrt{s}-\omega_k^{\mathrm{
  thr}}}{\sqrt{s}+\omega_k^{\mathrm{thr}}}\right|-i\pi\right)
  \mathcal{F}(\tilde{\mathrm{q}}_k)\right]
\label{eq:9}
\end{align}
with 
\begin{align}
\mathcal{F}(\mathrm{q})=\frac{\mathrm{q}}{2}\,
  V^{J(fk)}_{\boldsymbol\lambda'\boldsymbol\lambda_k}(\mathrm{p}',
  \mathrm{q})T^{J(ki)}_{\boldsymbol\lambda_k\boldsymbol\lambda}
  (\mathrm{q},\mathrm{p}). 
\label{eq:10}
\end{align}
Here $\omega_k=\omega_1^k+\omega_2^k$ and
$\omega_k^{\mathrm{thr}}=m_1^k+m_2^k$. For an open channel,
$\tilde{\mathrm q}_k$ is the on-shell momentum satisfying
$\omega_k=\sqrt{s}$. Since our aim here is to compute physical 
transition amplitudes, we do not analytically continue the on-shell
terms in Eq.~\eqref{eq:9} to closed channels. Their off-shell
contributions are evaluated using the regular integral in
Eq.~\eqref{eq:7}.

To obtain definite spin-parity ($J^P$) assignments, the helicity
amplitudes are transformed into the $LSJ$ basis. The relation between the
helicity and $LSJ$ bases is given by the unitary transformation
\begin{align}
|JM; LS\rangle &= \sum_{\lambda_1, \lambda_2} \sqrt{\frac{2L+1}{2J+1}} 
( L 0; S \lambda | J \lambda ) 
( s_1 \lambda_1; s_2 -\lambda_2 | S \lambda ) 
|JM; \lambda_1 \lambda_2\rangle.
\label{eq:11}
\end{align}
This basis transformation enables the separation of states with
distinct parity eigenvalues. Furthermore, once isospin
symmetry is imposed, the transition amplitudes can be decomposed
according to the $I(J^P)$ quantum numbers. In the present work, we
investigate the $I(J^P)=\frac12(0^\pm)$, $\frac12(1^\pm)$, and
$\frac12(2^\pm)$ sectors. A pole is found only in $1^+$
(Sec.~\ref{sec:3-1}), on which the detailed discussion below
concentrates. In this sector the
complete $LSJ$ space consists of $D^*D_s$ and $DD_s^*$ in
${}^3S_1,{}^3D_1$, and $D^*D_s^*$ in
${}^3S_1,{}^3D_1,{}^5D_1$. The helicity equation includes all these
components, even where only the ${}^3S_1$ amplitudes are displayed.

\subsection{Effective Lagrangian for heavy and light mesons}
\label{sec:2-2} 
The effective Lagrangians that describe the interactions of heavy
mesons with light pseudoscalar and vector mesons follow from the
spontaneous breaking of chiral
symmetry~\cite{Coleman:1969sm,Callan:1969sn} and from hidden local
symmetry~\cite{Bando:1984ej,Bando:1985rf}, respectively. At leading
order in the heavy-quark expansion, the pseudoscalar- and vector-meson
interactions can be organized as 
\begin{align}
\label{eq:12}
\mathcal{L}_{\mathrm{heavy}}^Q &= ig\,\mathrm{Tr}[H_b \gamma_\mu
  \gamma_5 \mathcal{A}_{ba}^\mu \bar{H}_a] 
  + i\beta\,\mathrm{Tr}[H_b v_\mu(\mathcal{V}-\rho)_{ba}^\mu 
  \bar{H}_a] 
  + i\lambda\,\mathrm{Tr}[H_b \sigma_{\mu\nu} F_{ba}^{\mu\nu} 
  \bar{H}_a].
\end{align}
The chiral currents are given by
$\mathcal{A}_\mu=(\xi^\dagger\partial_\mu\xi-\xi\partial_\mu\xi^\dagger)/2$  
and
$\mathcal{V}_\mu =
(\xi^\dagger\partial_\mu\xi+\xi\partial_\mu\xi^\dagger)/2$,  
where $\xi=\exp(i\mathcal M/f_\pi)$ is the coset field of the
pseudo-Nambu-Goldstone bosons. The field strength for the vector
mesons is defined by 
$F_{\mu\nu}=\partial_\mu\rho_\nu-\partial_\nu\rho_\mu+[\rho_\mu,\rho_\nu]$. 
The superfield in $4\times4$ Dirac space is
\begin{align}
\label{eq:13}
H_a = \frac{1+\slashed{v}}{2} \left[ \slashed{P}_a^* - P_a \gamma_5 \right],
\;\;\; \bar{H}_a = \gamma^0 H_a^\dagger \gamma^0.
\end{align}
This expression reflects HQSS and
satisfies the constraint $\slashed{v}H=H$. To absorb the $m_Q$
dependence, the pseudoscalar and vector meson fields are normalized as
\begin{align}
\label{eq:14}
\langle 0|P_a|(Q\bar{q})_a,0^-\rangle = \sqrt{M_H}, \quad
\langle 0|P_a^{*\mu}|(Q\bar{q})_a,1^-\rangle = \epsilon^\mu 
\sqrt{M_H},
\end{align}
where $M_H$ denotes the mass of the heavy meson, which tends to
infinity in the heavy-quark limit, and $\epsilon^\mu$ is the
polarization vector.
Expanding Eq.~\eqref{eq:12}, we group the interaction Lagrangians
according to the heavy-meson fields involved:
\begin{align}
\mathcal{L}_{PP^*\mathcal{M}} &= -\frac{2g}{f_\pi} P_a^{\dagger} 
  \partial^\mu \mathcal{M}_{ba} P_{b\mu}^* + \mathrm{H.c.}, \cr
\mathcal{L}_{P^*P^*\mathcal{M}} &= -\frac{2ig}{f_\pi} 
  \varepsilon^{\mu\nu\alpha\beta} P_{a\alpha}^{*\dagger} v_\beta 
  \partial_\mu \mathcal{M}_{ba} P_{b\nu}^*, 
\label{eq:15}
\end{align}
and
\begin{align}
\mathcal{L}_{PPV} &= -\sqrt{2} \beta g_V P_a^\dagger v_\mu 
  V_{ba}^\mu P_b, \cr
\mathcal{L}_{P^*P^*V} &= \sqrt{2} \beta g_V P_{a\nu}^{*\dagger} 
  v_\mu V_{ba}^\mu P_b^{*\nu} + 2\sqrt{2}i\lambda g_V 
  P_{a\mu}^{*\dagger} P_{b\nu}^* V_{ba}^{\mu\nu}, \cr
\mathcal{L}_{PP^*V} &= -2\sqrt{2}\lambda g_V 
  \varepsilon_{\beta\alpha\mu\nu} (P_a^\dagger v^\beta P_b^{*\alpha} 
  + P_a^{*\dagger\alpha} v^\beta P_b) (\partial^\mu V^\nu)_{ba},
\label{eq:16}
\end{align}
where $V^{\mu\nu}=\partial^\mu V^\nu-\partial^\nu V^\mu$ at the
one-vector-meson vertices. The matrices $\mathcal{M}$ and $V^\mu$
contain the pseudoscalar octet and the ideally mixed vector nonet,
respectively:
\begin{align}
\label{eq:17}
\mathcal{M} = \left(
\begin{array}{ccc}
\frac{\pi^0}{\sqrt{2}}+\frac{\eta}{\sqrt{6}} & \pi^+ & K^+
\\ 
\pi^- & -\frac{\pi^0}{\sqrt{2}}+\frac{\eta}{\sqrt{6}} & K^0
\\
K^- & \bar{K}^0 & -\sqrt{\frac{2}{3}}\eta
\end{array}\right), \;\;\;
\rho^\mu = i\frac{g_V}{\sqrt{2}} V^\mu,\;\;\;
V^\mu=\left(\begin{array}{ccc}
\frac{\rho^0}{\sqrt{2}}+\frac{\omega}{\sqrt{2}} & \rho^+ & K^{*+}
\\ 
\rho^- & -\frac{\rho^0}{\sqrt{2}}+\frac{\omega}{\sqrt{2}} & K^{*0}
\\
K^{*-} & \bar{K}^{*0} & \phi
\end{array}\right)^\mu.
\end{align}
The scalar interaction is introduced separately as a phenomenological
parametrization. For the scalar exchanges
$S=\sigma,f_0(980)$, we use
\begin{align}
\mathcal{L}_{PPS} = -2\sum_a g_{S P_aP_a}
P_a^\dagger P_a S,
\quad \mathcal{L}_{P^*P^*S} = 2\sum_a g_{S
P_a^*P_a^*} P_{a\mu}^{*\dagger} P_a^{*\mu} S.
\label{eq:18}  
\end{align}
To convert the velocity-normalized heavy fields to relativistic
meson fields $\phi_a$ and $\phi_a^{*\mu}$, we write
$P_a=\sqrt{m_{P_a}}\,\phi_a$ and
$P_a^{*\mu}=\sqrt{m_{P_a^*}}\,\phi_a^{*\mu}$.
Thus Eq.~\eqref{eq:18} generates scalar vertices proportional to
$-2g_{S PP}\sqrt{m_1m_2}$ and
$+2g_{S P^*P^*}\sqrt{m_1m_2}\,\epsilon_f^*\cdot\epsilon_i$,
respectively. The scalar couplings quoted below are dimensionless.
In the derivative interactions, the velocity is represented by
$v_\mu\to i\overleftrightarrow{\partial}_\mu/(2\sqrt{m_1m_2})$,
with $\overleftrightarrow{\partial}
=\overrightarrow{\partial}-\overleftarrow{\partial}$.
For incoming and outgoing waves $e^{-ip\cdot x}$ and
$e^{+ip'\cdot x}$, this gives
$v_\mu\to(p+p')_\mu/(2\sqrt{m_1m_2})$.

The coupling constants are determined as follows. The value $g=0.59$
is fixed by the $D^{*+} \to D^0 \pi^+$ decay width. We use
$f_\pi=132$~MeV, $g_V=5.8$, $\beta=0.9$, and
$\lambda=-0.56~\mathrm{GeV}^{-1}$, following the phenomenological
estimates discussed in Refs.~\cite{Casalbuoni:1996pg,Isola:2003fh}.
While $\sigma$ is often treated as the chiral partner of the pion, its
large width suggests that it is better modeled as a correlated $2\pi$
exchange in the $S$ wave. We accordingly take $g_{\sigma DD}=1.50$ and
$g_{\sigma D^*D^*}=5.21$ from such an exchange, as evaluated in
Ref.~\cite{Kim:2019rud}. To estimate the $\sigma$ couplings to strange
heavy mesons, we employ a 
flavor mixing of $\sigma$ meson without explicitly evaluating
correlated $K\bar{K}$ exchange. We introduce the normalized flavor
matrices $S_1=\mathrm{diag}(1,1,1)/\sqrt{3}$ and
$S_8=\mathrm{diag}(1,1,-2)/\sqrt{6}$ and write
\begin{align}
\sigma&=\cos\theta\,S_1+\sin\theta\,S_8, \label{eq:19}\\
f_0&=-\sin\theta\,S_1+\cos\theta\,S_8. \label{eq:20}
\end{align}
In the scalar-coupling subscripts below, $H=P,P^*$ labels an
individual heavy meson. Its couplings to the scalar flavor singlet and
octet are denoted by $g_{P^{(*)}}^1$ and $g_{P^{(*)}}^8$,
respectively. We assume $g_{P^{(*)}}^8/g_{P^{(*)}}^1=1$ and hence
write $g_{P^{(*)}}^1=g_{P^{(*)}}^8\equiv g_{P^{(*)}}$. We use the
singlet-octet mixing angle $\theta=19^\circ$ from
Ref.~\cite{Oller:2003scalar}. With $q=u,d$, the couplings are
\begin{align}
g_{\sigma H_qH_q}&=g_{P^{(*)}}\left(\frac{\cos\theta}{\sqrt3}
                   +\frac{\sin\theta}{\sqrt6}\right), &
g_{\sigma H_sH_s}&=g_{P^{(*)}}\left(\frac{\cos\theta}{\sqrt3}
                   -\frac{2\sin\theta}{\sqrt6}\right), \notag\\
g_{f_0H_qH_q}&=g_{P^{(*)}}\left(-\frac{\sin\theta}{\sqrt3}
                   +\frac{\cos\theta}{\sqrt6}\right), &
g_{f_0H_sH_s}&=g_{P^{(*)}}\left(-\frac{\sin\theta}{\sqrt3}
                   -\frac{2\cos\theta}{\sqrt6}\right).
\label{eq:21}
\end{align}
The nonstrange $\sigma$ coupling constants fix the common coupling as
\begin{align*}
g_{P^{(*)}}=\frac{g_{\sigma H_qH_q}}
{\cos\theta/\sqrt3+\sin\theta/\sqrt6},
\qquad g_P=2.210,\quad g_{P^*}=7.675.
\end{align*}
Numerically, in the order $(DD,D_sD_s,D^*D^*,D_s^*D_s^*)$, these give
\begin{align}
g_\sigma&=(1.500,\ 0.619,\ 5.210,\ 2.150),\notag\\
g_{f_0}&=(0.438,\ -2.121,\ 1.520,\ -7.368).
\label{eq:22}
\end{align}
We use the effective mass $m_\sigma=500$~MeV and take
$m_{f_0}=990$~MeV from the PDG mass estimate~\cite{PDG:2026scalar}.

\subsection{Construction of the kernel matrices}
\label{sec:2-3}

\begin{figure}[htp]
\centering
\includegraphics[scale=.8]{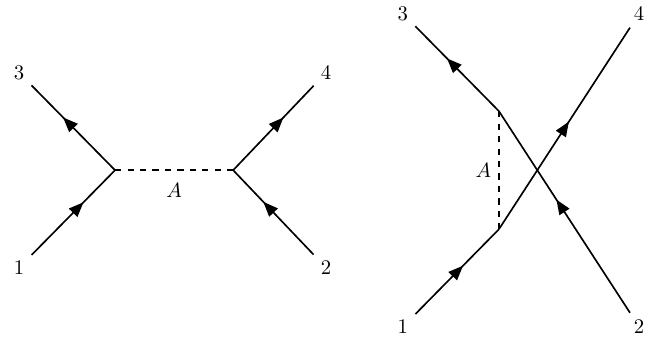}
\caption{$t$- and $u$-channel Feynman diagrams for the Born kernel of
the process $12\to34$, where $A$ denotes the exchanged meson.}
\label{fig:4}
\end{figure}

In the present work, we consider two-body hadronic channels with charm
number $C=+2$ and strangeness $\mathcal{S}=+1$ in the energy region near
the open-charm thresholds, which is directly relevant to the
$T_{cc\bar{s}}$ state. Invoking HQSS, which places
pseudoscalar and vector charmed mesons within a common spin doublet,
we include $D$, $D^*$, $D_s$, and $D_s^*$ in the channel space. This
leads to four distinct two-body hadronic channels: $DD_s$, $D^*D_s$,
$DD_s^*$, and $D^*D_s^*$, all carrying the total isospin $I=1/2$.
In each channel, particle 1 is the nonstrange $D^{(*)}$ and particle 2 is the
strange $D_s^{(*)}$. We use the same ordering when
extracting the pole residues. The interaction kernel is accordingly
constructed as a $4\times4$ matrix of transition amplitudes:
\begin{align}
\mathcal{V} = 
\begin{pmatrix}
\mathcal{V}_{DD_s\to DD_s} & \mathcal{V}_{D^*D_s\to DD_s} &
\mathcal{V}_{DD_s^*\to DD_s} & \mathcal{V}_{D^*D_s^*\to DD_s} \\ 
\mathcal{V}_{DD_s\to D^*D_s} & \mathcal{V}_{D^*D_s\to D^*D_s} &
\mathcal{V}_{DD_s^*\to D^*D_s} & \mathcal{V}_{D^*D_s^*\to D^*D_s} \\ 
\mathcal{V}_{DD_s\to DD_s^*} & \mathcal{V}_{D^*D_s\to DD_s^*} &
\mathcal{V}_{DD_s^*\to DD_s^*} & \mathcal{V}_{D^*D_s^*\to DD_s^*} \\ 
\mathcal{V}_{DD_s\to D^*D_s^*} & \mathcal{V}_{D^*D_s\to D^*D_s^*} &
\mathcal{V}_{DD_s^*\to D^*D_s^*} & \mathcal{V}_{D^*D_s^*\to D^*D_s^*} 
\end{pmatrix}.
\label{eq:23}
\end{align}
Rows label final channels and columns initial channels in the stated
ordering. Each element of the kernel matrix is
obtained by summing all Feynman invariant amplitudes generated by the
relevant interactions. Since our primary aim is to investigate the
dynamical generation of exotic states through coupled-channel
dynamics, we exclude $s$-channel pole contributions and consider only
$t$- and $u$-channel diagrams. The Born kernel contributions for the
process $12\to34$ depicted in Fig.~\ref{fig:4} are expressed as 
\begin{align}
\label{eq:24}
\mathcal{V}_{i\to j}^A= 
\mathcal{C}_I F_{A,1}F_{A,2} \Gamma_1^A \mathcal{P}^A
\Gamma_2^A,
\end{align}
where $A$ labels the exchanged particle. The vertex functions
$\Gamma^A$ are determined by the Lorentz structure of the effective
Lagrangians in Eqs.~\eqref{eq:15}, \eqref{eq:16}, and~\eqref{eq:18},
with the overall amplitude convention of Eq.~\eqref{eq:1}. The propagator
$\mathcal{P}^A$ for spin-0 and spin-1 exchange mesons is given by
\begin{align}
\mathcal{P}^A(q) &= \frac{1}{q^2 - m_A^2} , \;\;\;
\mathcal{P}^A_{\mu\nu}(q) = \frac{-g_{\mu\nu} + \frac{q_\mu q_\nu}{m_A^2}}
{\,q^2 - m_A^2\,},
\label{eq:25}
\end{align}
respectively. To ensure the convergence of the Born amplitudes at
large momentum transfer, phenomenological form factors $F_{A,r}$ are
introduced at each vertex $r=1,2$:
\begin{align}
\label{eq:26}
F_A(q^2) = \left(\frac{n\Lambda_A^2-m_A^2}
{n\Lambda_A^2+\bm{q}^2}\right)^{n},
\end{align}
where $q$ is the three-momentum transfer. We turn off the energy
dependence of the form factors. The exponent $n$ is chosen
according to the power of momentum in the vertex function, so that
each amplitude is suppressed appropriately at high momenta.
Table~\ref{tab:2} lists the values of $n$ used in the calculation. In
the limit $n\to\infty$, $F_A$ reduces to a Gaussian form. 

The cutoff masses are not fitted to data in this work. We parametrize
them as $\Lambda_A=m_A+\Lambda_0$, using one common reduced cutoff
$\Lambda_0$ for all exchanged particles. This prescription has been
used throughout our coupled-channel calculations 
of light axial-vector mesons, heavy-meson molecules, and hidden-charm
pentaquarks~\cite{Clymton:2022jmv,Clymton:2023txd,Clymton:2024pql,
Kim:2023htt,Kim:2025ado,Clymton:2024fbf,Clymton:2025hez,
Clymton:2025zer}.  Its theoretical ground was discussed in
Ref.~\cite{Kim:2025ado}. A parametrization of the same functional form,
$\Lambda_A=m_A+\eta\Lambda_{\mathrm{QCD}}$, has also been found useful
in the quite different context of final-state interactions in hadronic
$B$ decays~\cite{Cheng:2004ru}. The prescription reduces the number of
independently varied cutoff parameters but does not remove regulator
dependence. In the present work, we adopt $\Lambda_0=600$~MeV
as a phenomenological reference, and vary it between $500$ and
$700$~MeV to examine the theoretical uncertainty of the numerical
results.  The coefficient $\mathcal{C}_I$ in Eq.~\eqref{eq:24} stands
for the SU(3) Clebsch-Gordan and isospin factors of each exchange
diagram. The corresponding values and the form-factor exponents are
listed in Table~\ref{tab:2}.
\setlength{\tabcolsep}{8pt}
\renewcommand{\arraystretch}{1.5}
\begin{table*}[!t]
\centering
\caption{Leading light-meson exchanges, channel types, SU(3) and isospin
factors $\mathcal{C}_I$, and form-factor exponents $(n_1,n_2)$ for the
kernel amplitudes of the $C=+2$, $\mathcal{S}=+1$ coupled-channel system. For
$12\to34$, the exponents refer to vertices
$(13,24)$ in the $t$ channel and $(14,23)$ in the $u$ channel.
Scalar flavor coefficients are included in the vertex couplings.}
\label{tab:2}
\begin{tabular}{c|c|c|c|c}
\hline
\hline
Reaction & Exchange & Channel & $\mathcal{C}_I$ & $(n_1,n_2)$ \\
\hline
\multirow[t]{3}{*}{$DD_s\to DD_s$}
& $\sigma$  & $t$ & $1$  & $(1,1)$ \\
& $f_0(980)$ & $t$ & $1$ & $(1,1)$ \\
& $K^*$     & $u$ & $1$    & $(1,1)$ \\
$DD_s\to D^*D_s$  & $K^*$ & $u$ & $1$ & $(1,2)$ \\
$DD_s\to DD_s^*$  & $K^*$ & $u$ & $1$ & $(2,1)$ \\
\multirow[t]{3}{*}{$DD_s\to D^*D_s^*$}
& $\eta$    & $t$ & $-1/3$ & $(2,2)$ \\
& $K$       & $u$ & $1$    & $(1,1)$ \\
& $K^*$     & $u$ & $1$    & $(2,2)$ \\
\hline
\multirow[t]{4}{*}{$D^*D_s\to D^*D_s$}
& $\sigma$  & $t$ & $1$    & $(2,1)$ \\
& $f_0(980)$ & $t$ & $1$ & $(2,1)$ \\
& $K$       & $u$ & $1$    & $(1,1)$ \\
& $K^*$     & $u$ & $1$    & $(2,2)$ \\
\multirow[t]{2}{*}{$D^*D_s\to DD_s^*$}
& $\eta$    & $t$ & $-1/3$ & $(2,2)$ \\
& $K^*$     & $u$ & $1$    & $(2,1)$ \\
\multirow[t]{3}{*}{$D^*D_s\to D^*D_s^*$}
& $\eta$    & $t$ & $-1/3$ & $(3,2)$ \\
& $K$       & $u$ & $1$    & $(2,1)$ \\
& $K^*$     & $u$ & $1$    & $(2,2)$ \\
\hline
\multirow[t]{4}{*}{$DD_s^*\to DD_s^*$}
& $\sigma$  & $t$ & $1$    & $(1,2)$ \\
& $f_0(980)$ & $t$ & $1$ & $(1,2)$ \\
& $K$       & $u$ & $1$    & $(1,1)$ \\
& $K^*$     & $u$ & $1$    & $(2,2)$ \\
\multirow[t]{3}{*}{$DD_s^*\to D^*D_s^*$}
& $\eta$    & $t$ & $-1/3$ & $(2,3)$ \\
& $K$       & $u$ & $1$    & $(1,2)$ \\
& $K^*$     & $u$ & $1$    & $(2,2)$ \\
\hline
\multirow[t]{5}{*}{$D^*D_s^*\to D^*D_s^*$}
& $\sigma$  & $t$ & $1$    & $(2,2)$ \\
& $f_0(980)$ & $t$ & $1$ & $(2,2)$ \\
& $\eta$    & $t$ & $-1/3$ & $(3,3)$ \\
& $K$       & $u$ & $1$    & $(2,2)$ \\
& $K^*$     & $u$ & $1$    & $(2,2)$ \\
\hline
\hline
\end{tabular}
\end{table*}

\section{Results and discussion}
\label{sec:3}
We solve the coupled-channel equation for the $C=+2$, $\mathcal{S}=+1$ system
with the interaction specified in Sec.~\ref{sec:2}. At
$\Lambda_0=600$~MeV, we search for bound-state poles on the physical
sheet and for resonance poles in the $I=1/2$, $J^P=0^\pm, 1^\pm,
2^\pm$ partial waves. The $0^+$, $1^-$, and $2^+$ ones include
$DD_s$. All partial waves except $0^+$ contain $D^*D_s$ and $DD_s^*$,
whereas $D^*D_s^*$ is considered in every partial wave. For each
channel, we include all $LSJ$ components allowed by parity and angular
momentum, and we scan the complex-energy plane for poles of the
transition amplitudes, with $\operatorname{Re}\sqrt{s}$ ranging from
$3.5$ to $4.3$~GeV and $\operatorname{Im}\sqrt{s}$ ranging from $-100$
to $0$~MeV in every channel. We find a bound-state pole only in the
$1^+$ channel, at $3972.90$~MeV. No other bound-state or resonance
poles are seen within this energy region. 

To investigate the dynamical origin and properties of the $1^+$ bound
state, we first examine the role of channel coupling in
Sec.~\ref{sec:3-1} and determine the pole position and channel coupling
strengths in Sec.~\ref{sec:3-2}. We then discuss the cross sections near
the open-charm thresholds in Sec.~\ref{sec:3-3} and trace the evolution
of the pole with the reduced cutoff mass in Sec.~\ref{sec:3-4}.
Finally, in Sec.~\ref{sec:3-5}, we examine the flavor classification
of the state and its relation to the nonstrange $T_{cc}$ sector.

\subsection{Role of channel coupling in pole generation}
\label{sec:3-1}
\begin{figure}[htbp]
  \centering
  \includegraphics[width=0.6\textwidth]{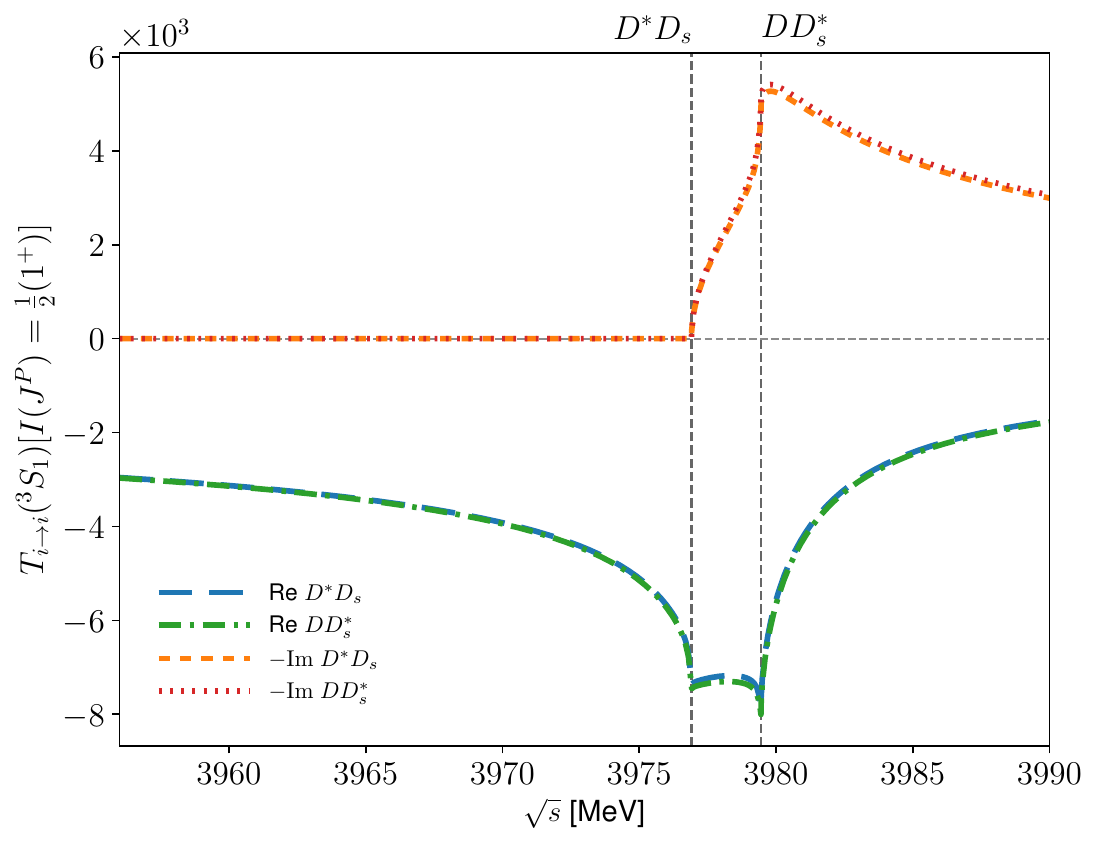}
  \caption{Elastic ${}^3S_1$ amplitudes with only $D^*D_s$ and
  $DD_s^*$ coupled at $\Lambda_0=600$~MeV. The vertical dashed lines
  mark the $D^*D_s$ and $DD_s^*$ thresholds. No bound-state pole is
  found in this subsystem.}
  \label{fig:5}
\end{figure}

\begin{figure}[htbp]
  \centering
  \includegraphics[width=0.485\textwidth]{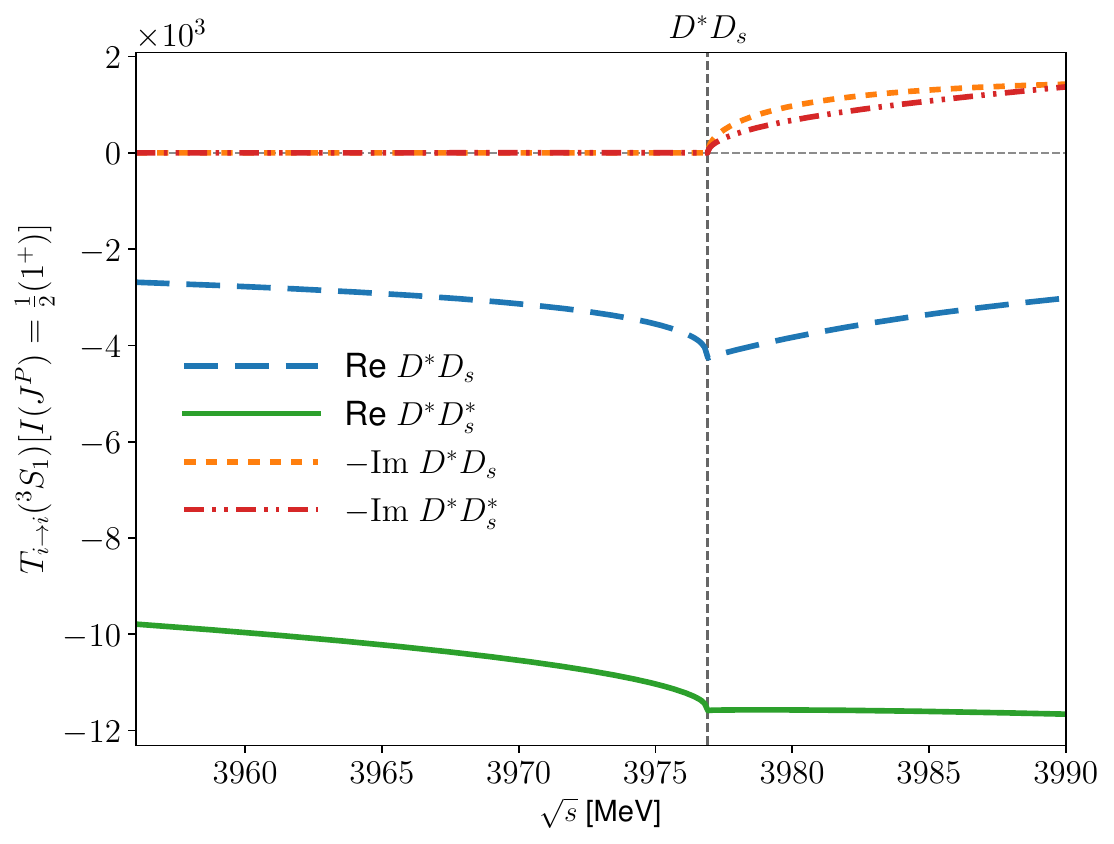}
  \includegraphics[width=0.485\textwidth]{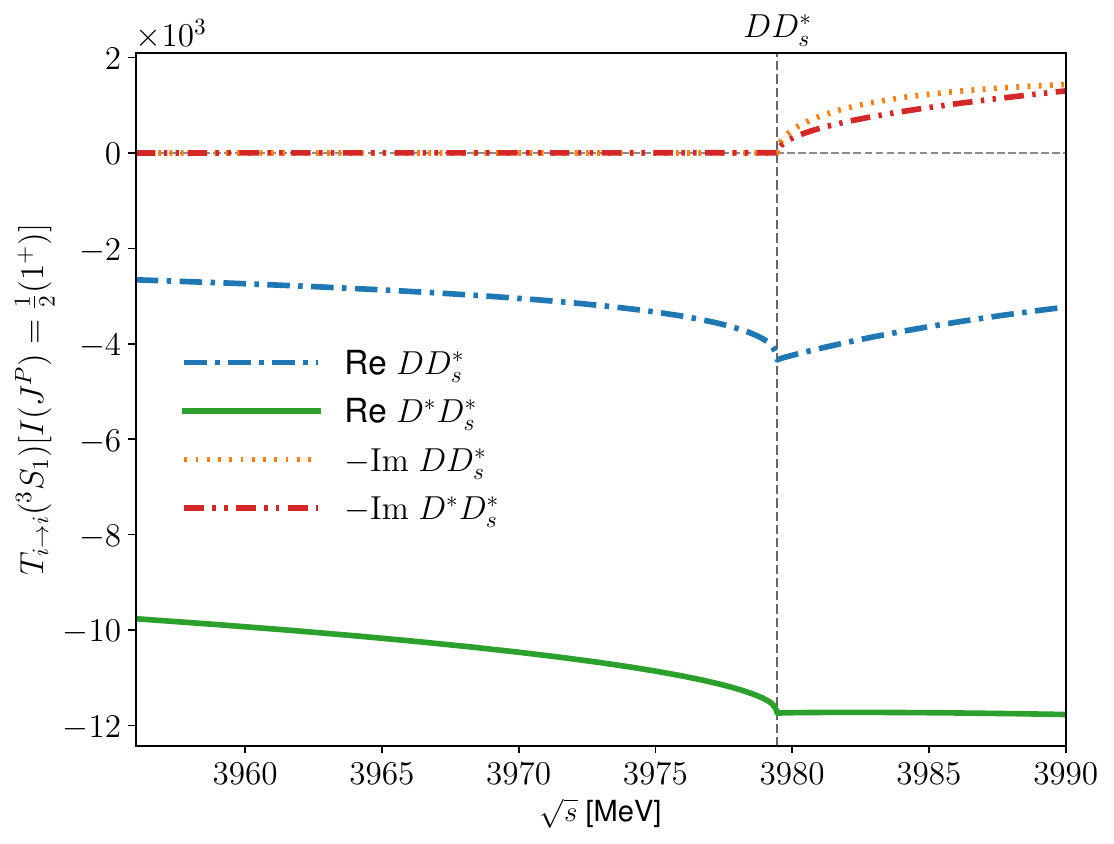}
  \caption{Elastic ${}^3S_1$ amplitudes for the
  $D^*D_s$--$D^*D_s^*$ (left) and $DD_s^*$--$D^*D_s^*$ (right)
  two-channel subsystems at $\Lambda_0=600$~MeV. The interaction
  parameters and amplitude conventions are the same as in
  Fig.~\ref{fig:5}. The vertical dashed lines mark the $D^*D_s$
  and $DD_s^*$ thresholds in the left and right panels, respectively.
  Neither subsystem produces the bound-state pole found in the full
  three-channel calculation.}
  \label{fig:6}
\end{figure}

At $\Lambda_0=600$~MeV, the full $I(J^P)=\frac1{2}(1^+)$ system has a
real subthreshold pole at $3972.90$~MeV, or $4.01$~MeV below the
$D^*D_s$ threshold. The $DD_s$ channel is excluded from this sector
by parity. To identify the roles of the remaining
three channels, we repeat the calculation for each pair of them.
Figures~\ref{fig:5} and~\ref{fig:6} show the resulting elastic
amplitudes for the three two-channel subsystems near the $D^*D_s$
and $DD_s^*$ thresholds. We focus on the elastic $S$-wave transition
amplitudes, which exhibit the clearest peaks or threshold structures.

Figure~\ref{fig:5} shows the ${}^3S_1$ elastic amplitudes near the
$D^*D_s$ and $DD_s^*$ thresholds when only these two channels are
coupled. The two elastic amplitudes are nearly identical because the
channels have similar masses and meson-exchange interactions as shown
in Tables~\ref{tab:1} and~\ref{tab:2}. The real parts of the transition
amplitudes remain negative throughout the $3.95$--$3.99$~GeV region
and exhibit cusps at both thresholds due to the coupling between the
channels. As the $D^*D_s$ channel opens, the magnitudes of the imaginary
parts rise rapidly and begin to decrease as the $DD_s^*$ channel
opens. Despite these threshold structures, no poles are found in our
pole search. 

We then examine the elastic amplitudes for the $D^*D_s$--$D^*D_s^*$
and $DD_s^*$--$D^*D_s^*$ subsystems using the same input parameters
and reduced cutoff masses. The results are shown in Fig.~\ref{fig:6}.
As before, the similarity between the $D^*D_s$ and $DD_s^*$ channels
leads to similar transition amplitudes in the two subsystems near
their respective thresholds. The $D^*D_s^*$ channel is closed in this
energy region, as its threshold lies about $140$~MeV above the
$D^*D_s$ and $DD_s^*$ thresholds. Nevertheless, its elastic amplitudes
have large magnitudes even near these lower thresholds. The threshold cusps
are weaker than those in the $D^*D_s$--$DD_s^*$
subsystem because the coupling between $D^*D_s$ and $DD_s^*$ is
absent. No poles are found in either two-channel subsystem.

\begin{figure}[htbp]
  \centering
  \includegraphics[width=0.6\textwidth]{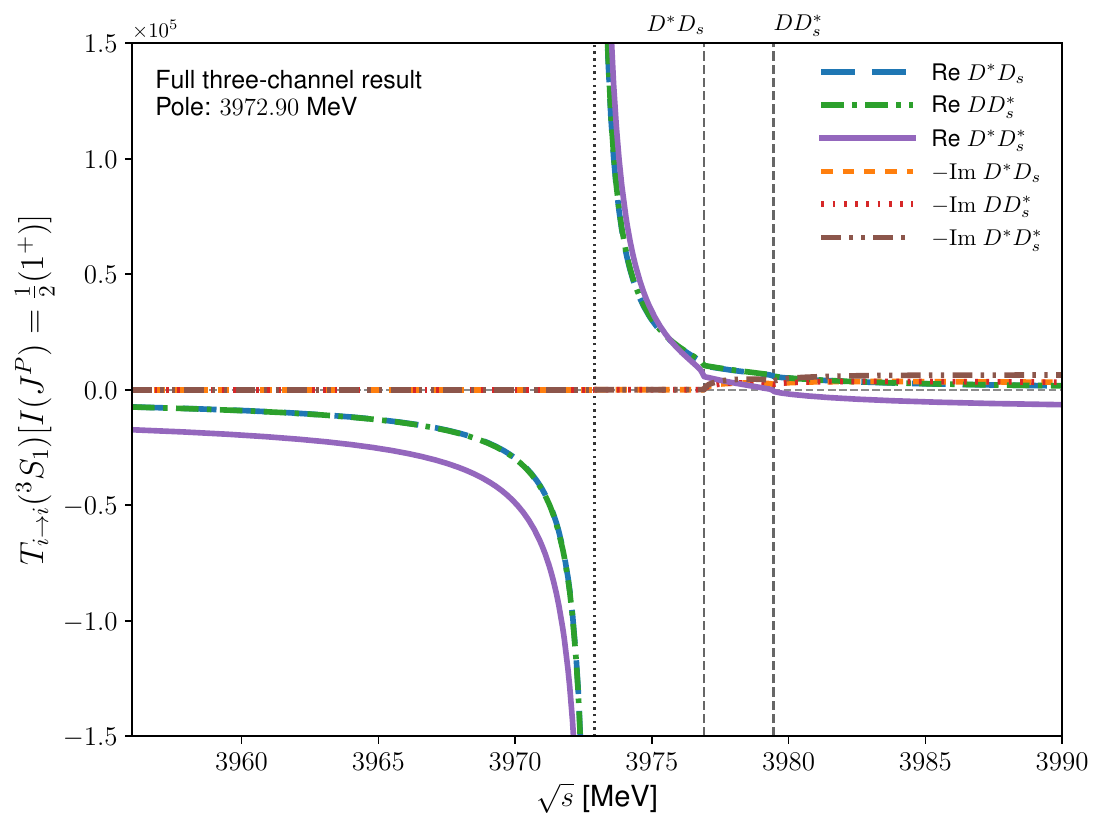}
  \caption{Elastic ${}^3S_1$ amplitudes for the full three-channel
  system with $D^*D_s$, $DD_s^*$, and $D^*D_s^*$ coupled at
  $\Lambda_0=600$~MeV. The interaction parameters and amplitude
  conventions are   the same as in Figs.~\ref{fig:5} and~\ref{fig:6}. 
  The dotted line marks the common pole at $3972.90$~MeV.
  The vertical dashed lines mark the $D^*D_s$ and $DD_s^*$ thresholds.
  }
  \label{fig:7}
\end{figure}

We finally couple all three channels in the $J^P=1^+$ wave.
Figure~\ref{fig:7} shows the resulting elastic transition amplitudes
in the same energy region as the two-channel results. Their real
parts exhibit a pole on the real energy axis below the lowest
threshold, $D^*D_s$. These comparisons show that coupling to
$D^*D_s^*$ provides the additional attraction needed to form this
bound state. We examine its precise pole position and channel
couplings in Sec.~\ref{sec:3-2}.

\subsection{Pole position and channel coupling strengths}
\label{sec:3-2}
To characterize the $T_{cc\bar{s}}$ bound state quantitatively, we
determine its pole position and coupling strengths to the hadronic
channels. The pole is located at $\sqrt{s_R}=3972.90$~MeV on the real
$\sqrt{s}$ axis, corresponding to a binding energy of $4.01$~MeV
relative to the $D^*D_s$ threshold. In the present calculation, we
treat the constituent mesons as stable and focus on the dynamical
generation of the pole through two-body scattering. Since the pole
lies below the lowest allowed two-body threshold, it represents a
bound state. A more complete treatment could include decays of the
constituent mesons, which can give the state a finite width through
kinematically allowed $DD_s\pi$ channels, such as $D^0D_s^+\pi^0$. The
possible role of $DD\pi$ channels was also discussed in our earlier
$T_{cc}$ study~\cite{Kim:2025ado}. These three-body channels and the
radiative $DD_s\gamma$ channels are not included in the present
calculation.

We now extract the coupling strengths $g_i$ from the transition
amplitudes near the pole~\cite{Clymton:2022jmv, Clymton:2023txd,
Kim:2023htt}:
\begin{align}
T_{i\to j}(\sqrt{s}) = 4\pi\frac{g_i g_j}{s - s_R} + T^{\mathrm{reg}}_{ij}(s),
\label{eq:27}
\end{align}
where $T^{\mathrm{reg}}_{ij}(s)$ denotes the part that is regular at
the pole. The indices $i$ and $j$ specify the hadronic channels and
their partial-wave components. We evaluate the residues near
$\sqrt{s_R}$, using the same numerical prescription as in the pole
search. The diagonal and off-diagonal residues determine the coupling
magnitudes and relative signs, respectively. In the adopted channel
basis, we fix the overall sign by choosing $g_{D^*D_s({}^3S_1)}$ to be
positive. The resulting pole position and coupling strengths are
listed in Table~\ref{tab:3}.
\setlength{\tabcolsep}{12pt}
\renewcommand{\arraystretch}{1.5}
\begin{table}[htp]
\centering
\caption{Coupling strengths $g_i$ (in GeV) for the
$I(J^P)=\frac1{2}(1^+)$ bound state at $\sqrt{s_R}=3972.90$~MeV. The
couplings are extracted from the residues in Eq.~\eqref{eq:27}, with
the overall sign fixed by $g_{D^*D_s({}^3S_1)}>0$. We list the
spin-triplet $S$- and $D$-wave components.}
\begin{tabular}{c|c}
\hline\hline
$\sqrt{s_R}$ & $3972.90$~MeV \\
\hline
$g_{D^*D_s({}^3S_1)}$ & $6.848$ \\
$g_{DD_s^*({}^3S_1)}$ & $-6.880$ \\
$g_{D^*D_s^*({}^3S_1)}$ & $8.083$ \\
$g_{D^*D_s({}^3D_1)}$ & $3.226\times10^{-3}$ \\
$g_{DD_s^*({}^3D_1)}$ & $-5.034\times10^{-3}$ \\
$g_{D^*D_s^*({}^3D_1)}$ & $-1.402\times10^{-1}$ \\
\hline\hline
\end{tabular}
\label{tab:3}
\end{table}

The most significant feature is that the two pseudoscalar-vector (PV)
${}^3S_1$ couplings have opposite signs and nearly equal magnitudes.
These couplings support a predominantly antisymmetric $\bm{3}_A$
flavor assignment, as discussed in Sec.~\ref{sec:3-5}. The large
${}^3S_1$ coupling to $D^*D_s^*$ indeed indicates a significant
coupled-channel effect involving this channel. The comparison between
the two-channel results in Figs.~\ref{fig:5} and~\ref{fig:6} and the
full solution in Fig.~\ref{fig:7} clearly confirms the role of this
channel in binding. The ${}^3D_1$ couplings are considerably smaller
in magnitude than the corresponding ${}^3S_1$ couplings, particularly
for the two PV channels. The coupling $g_{D^*D_s^*({}^3D_1)}$ also has
the opposite sign to the corresponding $S$-wave coupling.

Note that the pole is dynamically generated solely by $t$- and
$u$-channel exchange diagrams without an explicit $s$-channel state.
Taking the above discussion into account, we may thus interpret it
as a molecular $T_{cc\bar{s}}$.

\subsection{Cross sections near the open-charm thresholds}
\label{sec:3-3}
Having determined the pole position and channel coupling strengths, we
now examine the cross sections near the open-charm thresholds. We
define the positive-parity cross sections, using the
same overall normalization for all reactions:
\begin{align}
\sigma^{+}_{i\to j}(s)
 &= \sum_{J}\sigma^{J,+}_{i\to j}(s)
  = \frac{1}{256\pi^3}\frac{1}{s}\frac{\mathrm p'}{\mathrm p}
    \frac1{(2s_{i1}+1)(2s_{i2}+1)} \sum_J (2J+1)
    \sum_{\bm{\lambda}',\bm{\lambda}}
    \left|
    T^{J,+}_{i\to j,\,
    \bm{\lambda}'\bm{\lambda}}
    (\mathrm p',\mathrm p;s)
    \right|^2,
\label{eq:28}
\end{align}
where $\mathrm p$ ($\mathrm p'$) denotes the CM momentum in the
initial (final) channel. The spins $s_{i1}$ and $s_{i2}$ of the
initial particles determine the averaging factor
$1/[(2s_{i1}+1)(2s_{i2}+1)]$. The helicity pairs
$\bm{\lambda}=(\lambda_1,\lambda_2)$ and
$\bm{\lambda}'=(\lambda_1',\lambda_2')$ label the initial and final
states, respectively. The amplitude $T^{J,+}$ denotes the
positive-parity projection of $T^J$ in the full helicity basis. The
factor $1/(256\pi^3)$ follows from the state normalization and
partial-wave projection in Eqs.~\eqref{eq:1} and~\eqref{eq:8}.

\begin{figure}[htbp]
  \centering
  \includegraphics[width=0.8\textwidth]{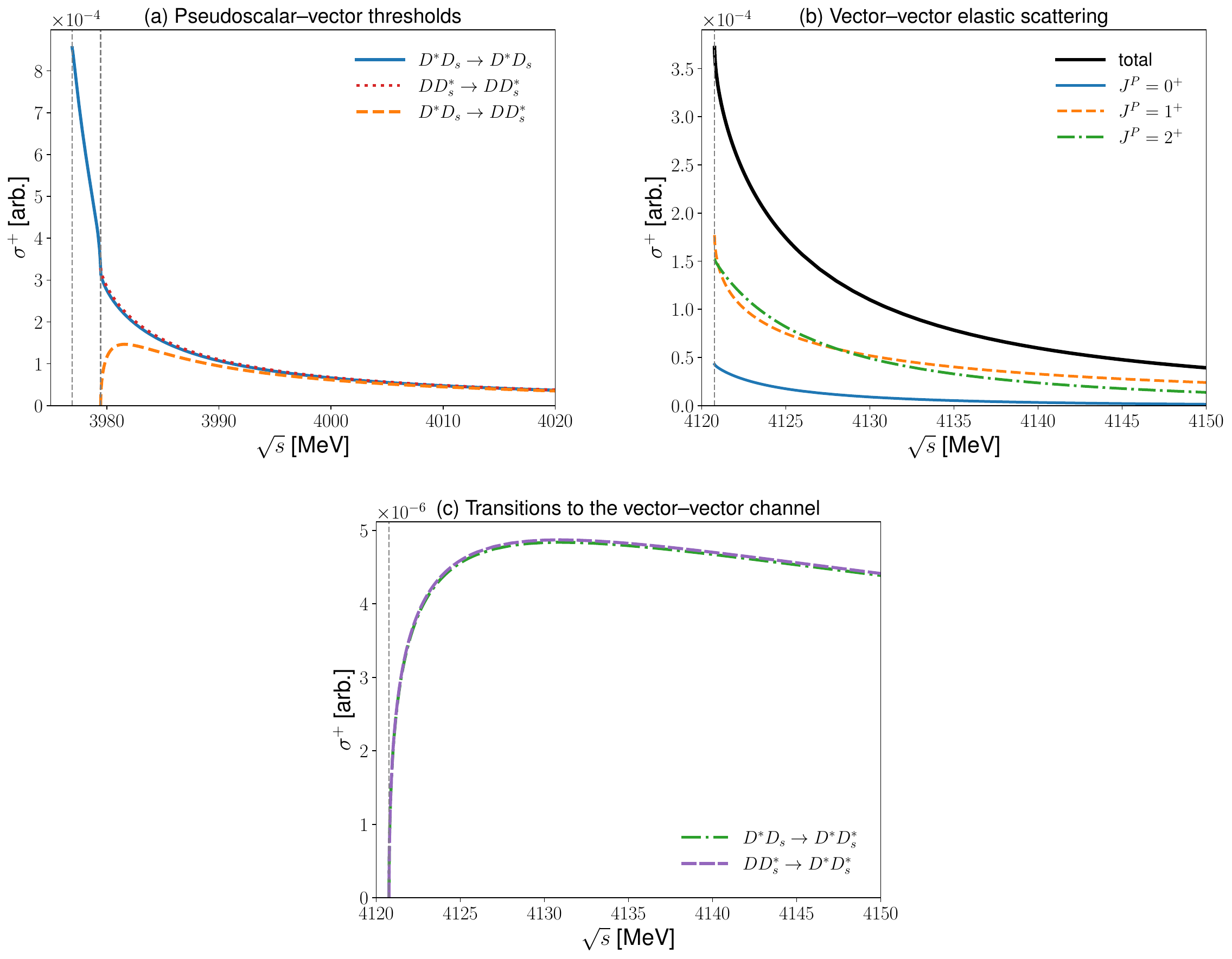}
  \caption{Open-charm elastic and transition cross sections in the
  $I=\frac12$ positive-parity sector at $\Lambda_0=600$~MeV, summed
  over $J$ and shown in arbitrary units. 
  (a) Elastic scattering and   the transition between the two
  pseudoscalar-vector channels near their thresholds. 
  (b) $D^*D_s^*$ elastic scattering. The black solid
  curve denotes the total $\sigma^+$. The other curves show the
  individual $J^P=0^+,1^+,2^+$ contributions. 
  (c) Transitions from $D^*D_s$ and $DD_s^*$ to $D^*D_s^*$. All cross
  sections include the initial spin average and use the same overall
  normalization. Gray dashed lines mark the relevant channel thresholds.}
  \label{fig:8}
\end{figure}

Figure~\ref{fig:8} shows the resulting cross sections at
$\Lambda_0=600$~MeV for energies above the thresholds of both the
initial and final channels. In panel (a), we present elastic
scattering and the transition between the two PV channels. The
$D^*D_s$ elastic cross section shows a pronounced threshold
enhancement and decreases with increasing energy. This enhancement
reflects the influence of the bound state pole identified in
Fig.~\ref{fig:7}, which lies below the $D^*D_s$ threshold. Above the
$DD_s^*$ threshold, the two elastic cross sections decrease with
increasing energy along the tail of this enhancement. The cross
section for the transition between the two PV channels starts from
zero at the $DD_s^*$ threshold, where the final-state momentum
vanishes, and approaches the elastic cross sections as the energy
increases. This behavior is expected from the physical similarity of
the two channels, as discussed above.

\begin{figure}[htbp]
  \centering
  \includegraphics[width=1.0\textwidth]{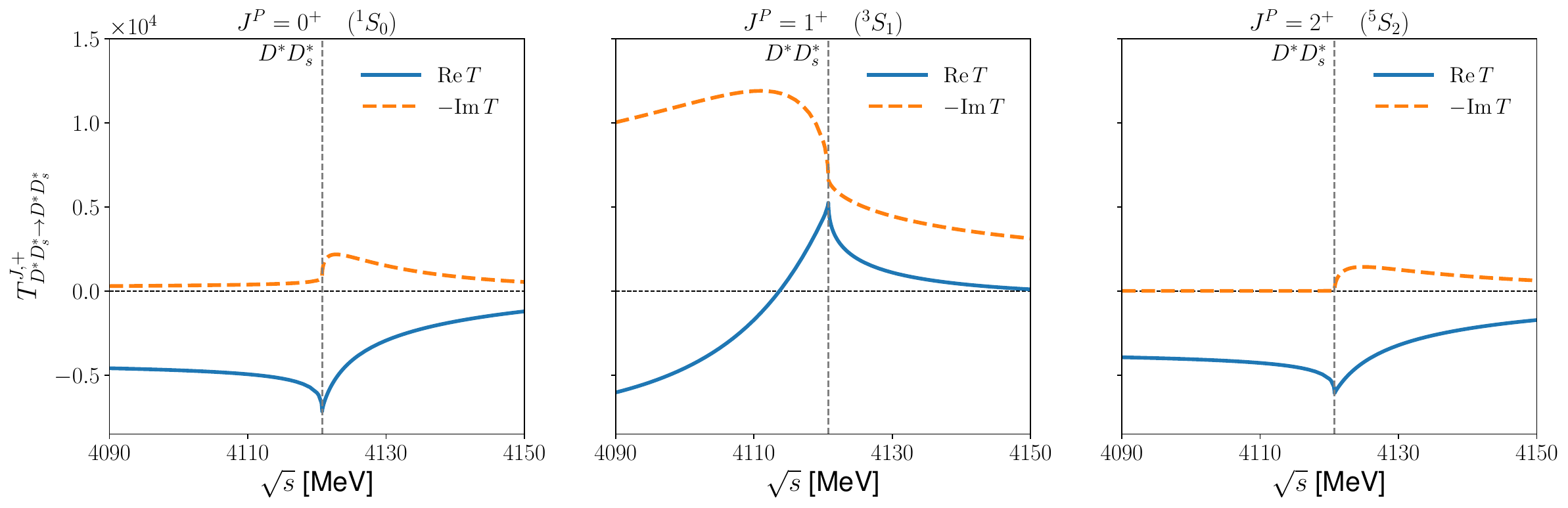}
  \caption{Diagonal $D^*D_s^*$ $S$-wave amplitudes near the
  $D^*D_s^*$ threshold at $\Lambda_0=600$~MeV. The panels show the
  ${}^1S_0$, ${}^3S_1$, and ${}^5S_2$ amplitudes from left to right.
  Solid and dashed curves denote $\mathrm{Re}\,T$ and
  $-\mathrm{Im}\,T$, respectively. Below threshold, the external
  relative momenta are set to zero.}
  \label{fig:9}
\end{figure}
The $D^*D_s^*$ elastic cross section is depicted in panel (b). The
black curve denotes the total positive-parity cross section. The
cross section is strongly enhanced at the $D^*D_s^*$ threshold and
decreases rapidly above it. Unlike the PV cases, the $J=2$
contribution is comparable to that of $J=1$, and the $J=0$ component
also contributes. To examine the amplitude structure
underlying this enhancement, we consider the $D^*D_s^*$ elastic
amplitudes in the ${}^1S_0$, ${}^3S_1$, and ${}^5S_2$ partial waves
near the threshold in Fig.~\ref{fig:9}. These amplitudes exhibit cusp
structures at the $D^*D_s^*$ threshold rather than the real-axis pole
behavior below the $D^*D_s$ threshold shown in Fig.~\ref{fig:7}.

Figure~\ref{fig:8}(c) shows the cross sections for transitions from the
PV channels to the vector-vector (VV) channel. The two transition
cross sections have even more similar line shapes than the PV elastic
cross sections because the small difference between the two PV
thresholds has a negligible kinematic effect at the $D^*D_s^*$ threshold,
which lies about $140$~MeV above them. 

\subsection{Dependence on the reduced cutoff mass}
\label{sec:3-4}
Since the reduced cutoff mass $\Lambda_0$ is the main cause for
theoretical uncertainties of the present work, we need to examine the
sensitivity of the pole position to it by varying its value between
$500$ and $700$~MeV, while holding the other interaction parameters
fixed. At each cutoff, we locate the pole from the homogeneous
coupled-channel BbS equation. Table~\ref{tab:4} lists the resulting
bound-state pole positions. The binding energy increases from
$4$~MeV to $34$ MeV as $\Lambda_0$ varies from $600$~MeV to
$700$~MeV. As the cutoff mass decreases, the pole position approaches
the $D^*D_s$ threshold. As a result, the bound pole disappears when
the value of $\Lambda_0$ is reduced below $550$~MeV, as shown in
Table~\ref{tab:4}. 
\setlength{\tabcolsep}{12pt}
\renewcommand{\arraystretch}{1.5}
\begin{table}[htp]
\centering
\caption{Dependence of the $T_{cc\bar{s}}$ bound-state pole position on
 the reduced cutoff mass $\Lambda_0=\Lambda_A-m_A$, where $A$ denotes
 the exchanged particle.
 $B=m_{D^*}+m_{D_s}-\sqrt{s_R}$ denotes the binding energy.
 All other interaction parameters are held fixed.
 Dashes indicate that no bound pole was found for
 $3500\leq\sqrt{s}<3976.91$~MeV.}
\begin{tabular}{c|c|c}
\hline\hline
$\Lambda_0$ [MeV] & $\sqrt{s_R}$ [MeV] & $B$ [MeV] \\
\hline
$500$ & --- & --- \\
$550$ & --- & --- \\
$600$ & $3972.90$ & $4.01$ \\
$650$ & $3961.20$ & $15.71$ \\
$700$ & $3942.87$ & $34.04$ \\
\hline\hline
\end{tabular}
\label{tab:4}
\end{table}

However, we have to verify whether the bound state pole evolves into a
virtual state as the reduced cutoff mass decreases.
To do that, we need to make an analytic continuation of the 
amplitude across the $D^*D_s$ cut, the lowest cut for this pole. We
trace the moving of the pole position as $\Lambda_0$ varies from
$600$ MeV down to $550$~MeV. We will call the physical sheet as sheet~I,
and the unphysical sheet, which arises when the pole position crosses
the cut, as sheet~II. Below the threshold, the on-shell relative
momentum in this channel is $q=+i\kappa$ on sheet~I and $q=-i\kappa$
on sheet~II, with $\kappa>0$.  

\begin{figure}[htbp]
  \centering
  \includegraphics[width=0.98\textwidth]{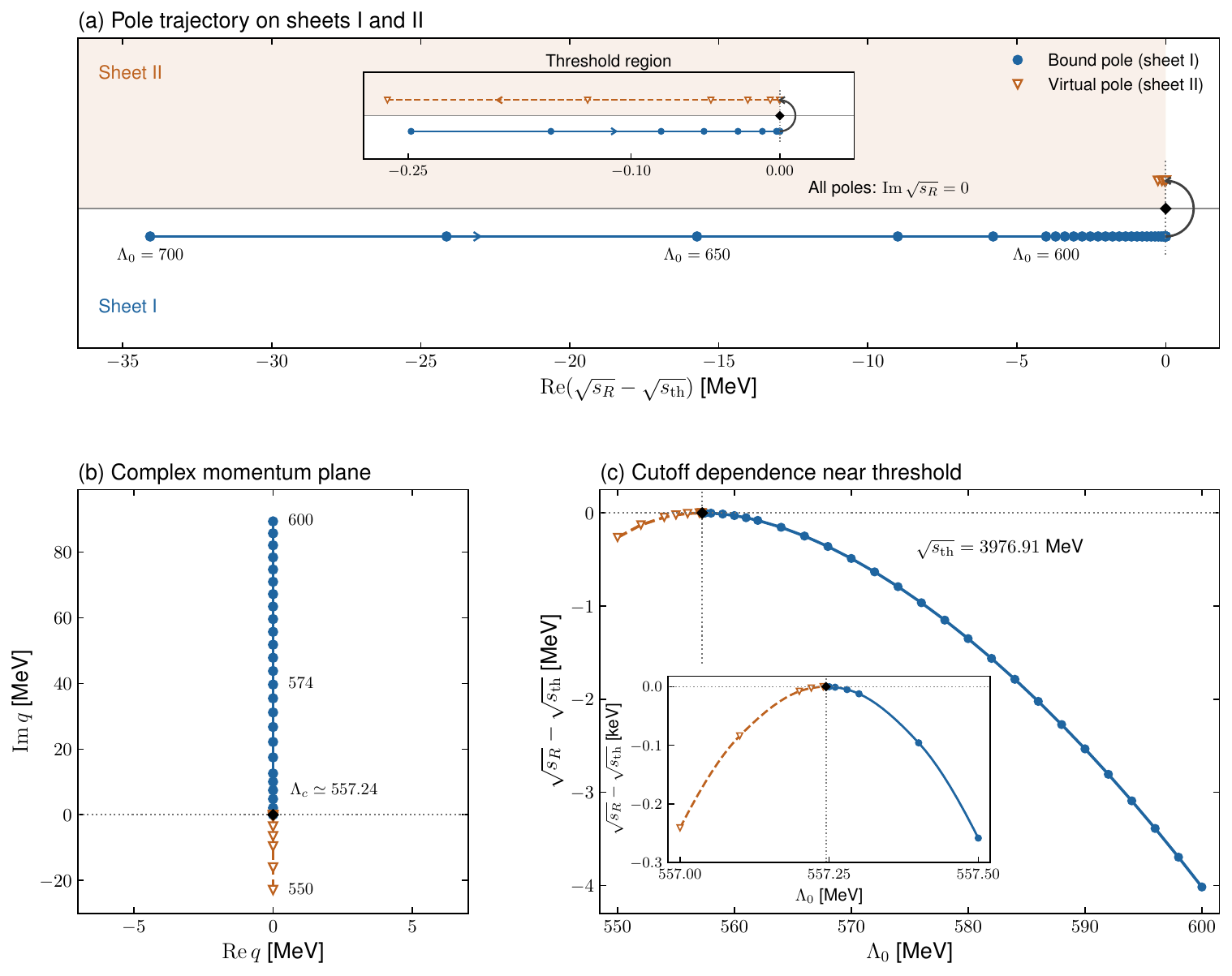}
  \caption{Cutoff mass dependence of the $1^+$ pole.
  (a) Pole energy relative to the $D^*D_s$ threshold,
  $\sqrt{s_{\rm th}}=3976.91$~MeV. The vertical separation of the
  sheets and the arrows across the threshold show schematically how
  the pole position crosses the cut. All pole energies are real.
  (b) Pole trajectory in the complex relative-momentum $q$ plane
  for $550\leq\Lambda_0\leq600$~MeV.
  (c) Pole energy as a function of $\Lambda_0$.
  Filled circles and open triangles denote bound poles on sheet~I
  and virtual poles on sheet~II, respectively. Black diamonds mark
  the threshold crossing at $\Lambda_c\simeq557.24$~MeV.
  Cutoff labels are in MeV.}
  \label{fig:10}
\end{figure}

Figure~\ref{fig:10} shows the pole trajectories in the energy and
momentum planes in panels (a) and (b), respectively. Panel (c)
displays the pole energy relative to the $D^*D_s$ threshold as a
function of $\Lambda_0$. As shown in panel (a), As $\Lambda_0$
decreases, the bound-state pole reaches the threshold 
at $\Lambda_0 \simeq 557.24~\mathrm{MeV}$ and crosses the cut onto
sheet~II, emerging as a virtual-state pole. At $\Lambda_0=550$~MeV,
the virtual pole lies about $0.264$~MeV below the 
$D^*D_s$ threshold. Panel (b) demonstrates this transition from the
positive to the negative imaginary $q$ axis through the origin, while
panel (c) shows the corresponding cutoff dependence of the pole
energy.
As mentioned previously, the value of the reduced cutoff mass
$\Lambda_0$ was mostly selected to be 600--700~MeV in the former 
investigations on various two-body processes for the hidden-charm
tetraquarks and pentaquarks~\cite{Kim:2023htt, Clymton:2024fbf,
  Clymton:2025zer, Kim:2025ado, Clymton:2025hez}, and smaller values
of $\Lambda_0$ than $\Lambda_0 =550$ MeV were barely
considered. Nevertheless, it is of great significance to examine how
the position of the bound-state pole is sensitive to the value of
$\Lambda_0$, as shown in Fig.~\ref{fig:10}. If the bound-state pole
exists, the minimum value of $\Lambda_0$ may possibly be around
600~MeV. We anticipate that the future experimental data will
constrain the $\Lambda_0$ value. 

\subsection{Flavor classification and relation to the nonstrange
  sector} 
\label{sec:3-5}
It is also very interesting to examine the flavor classification of
the $cc\bar{q}\bar{q}$ tetraquarks within the molecular picture.
A doubly charmed tetraquark can then be regarded as a system of two 
color-singlet mesons whose interaction is primarily mediated by meson
exchange, with its flavor structure constrained by SU(3) symmetry.
Since the low-lying $c\bar{q}$ mesons belong to the $\bar{\bm{3}}$
representation of flavor SU(3) symmetry, they can be decomposed into  
irreducible representations as
$\bar{\bm{3}}\otimes\bar{\bm{3}}=\bar{\bm{6}}_S\oplus\bm{3}_A$. 

In the nonstrange sector, the light-flavor wave functions of the
isoscalar $T_{cc}^{I=0}$ and isovector $T_{cc}^{I=1}$ are $(\bar u\bar
d-\bar d\bar u)/\sqrt{2}$ and $(\bar u\bar d+\bar d\bar u)/\sqrt{2}$,
respectively. The former belongs to $\mathbf{3}_A$ and the latter to
$\bar{\mathbf{6}}_S$. Both representations contain an
$\mathcal{S}=+1$, $I=1/2$ multiplet that accommodates $T_{cc\bar s}$,
as illustrated in Fig.~\ref{fig:1}. To determine the flavor assignment
of our bound state, we consider its dominant ${}^3S_1$ components
listed in Table~\ref{tab:3} and examine their behavior under the
exchange $\bar q\leftrightarrow\bar s$. Under this exchange, $D^*D_s$
becomes $D_s^*D$. Expressing the resulting state in our convention, in
which particle~1 is the nonstrange meson, requires interchanging the
meson positions and coupling their spins in the reversed order. The
orbital and spin parts then yield the factors $(-1)^L$ and
$(-1)^{s_1+s_2-S}$, respectively, where $s_1$ and $s_2$ denote the
meson spins and $S$ their total spin, respectively. With $L=0$ and
$S=1$, the combined factor is $+1$ for the $D^*D_s$ and $DD_s^*$
channels, while it is $-1$ for $D^*D_s^*$. The flavor exchange thus
maps $D^*D_s$ onto $DD_s^*$ and changes the sign of
$D^*D_s^*({}^3S_1)$. It follows that the combination
$(D^*D_s-DD_s^*)/\sqrt{2}$ and the $D^*D_s^*({}^3S_1)$ 
channel transform as $\mathbf{3}_A$, whereas
$(D^*D_s+DD_s^*)/\sqrt{2}$ transforms as $\bar{\mathbf{6}}_S$. 

We now use this classification to identify the dominant flavor
component of the bound state from its pole residues. Table~\ref{tab:3}
shows that the ${}^3S_1$ couplings to $D^*D_s$ and $DD_s^*$ are nearly
equal in magnitude but opposite in sign. Their difference therefore
enhances the coupling to $(D^*D_s-DD_s^*)/\sqrt{2}$, while their sum
largely cancels in the coupling to $(D^*D_s+DD_s^*)/\sqrt{2}$. The
pole thus couples strongly to the $\bm{3}_A$ combination, while it
couples only weakly to the $\bar{\bm{6}}_S$ combination. Its sizable
coupling to $D^*D_s^*({}^3S_1)$, which also transforms as $\bm{3}_A$,
supports the same assignment. Taken together, the $S$-wave pole
residues indicate that the bound state belongs predominantly to
$\bm{3}_A$.

\setlength{\tabcolsep}{10pt}
\renewcommand{\arraystretch}{1.5}
\begin{table}[htp]
\centering
\caption{Five isospin multiplets of the $cc\bar q\bar q$ system
and their $J^P=1^+$ pole results. The zero strangeness ($\mathcal{S}=0$) states
come from Ref.~\cite{Kim:2025ado}. The $\mathcal{S}=+1$ states are taken from
this work at $\Lambda_0=600$~MeV. The $\mathcal{S}=+2$ channel has not been
calculated. ``Predicted'' and ``generated'' both refer to model poles.}
\begin{tabular}{l|c|c|l}
\hline\hline
State & Rep. & $(\mathcal{S},\,I)$ & Status \\
\hline
$T_{cc}^{I=0}$   & $\bm{3}_A$       & $(0,\;0)$
  & predicted near $3823$~MeV~\cite{Kim:2025ado} \\
$T_{cc}^{I=1}$   & $\bar{\bm{6}}_S$ & $(0,\;1)$
  & predicted near $3875$~MeV~\cite{Kim:2025ado} \\
$T_{cc\bar{s}}$  & $\bm{3}_A$       & $(+1,\;\tfrac{1}{2})$
  & generated (this work) \\
$T_{cc\bar{s}}$  & $\bar{\bm{6}}_S$ & $(+1,\;\tfrac{1}{2})$
  & no additional pole found (this work) \\
$T_{cc\bar{s}\bar{s}}$ & $\bar{\bm{6}}_S$ & $(+2,\;0)$
  & not calculated \\
\hline\hline
\end{tabular}
\label{tab:5}
\end{table}
The dominance of the $\bm{3}_A$ component favors identifying the
present $J^P=1^+$ bound state as the strange partner of the isoscalar
$T_{cc}$ predicted in Ref.~~\cite{Kim:2025ado}. Table~\ref{tab:5}
summarizes this relation within the five isospin multiplets of the
$cc\bar q\bar q$ system. In the nonstrange sector,
Ref.~~\cite{Kim:2025ado} predicts an isoscalar $\bm{3}_A$ state near
$3823$~MeV and an isovector $\bar{\bm{6}}_S$ state near $3875$~MeV. In
the singly strange sector, our calculation generates a bound state
dominated by $\bm{3}_A$, while no additional pole associated with
$\bar{\bm{6}}_S$ is found with the same interaction parameters. Thus,
although both representations contain an $\mathcal{S}=+1$, $I=1/2$
multiplet, the coupled-channel dynamics produces only one bound state
in the present calculation. The two $\bm{3}_A$ states also share a
similar coupled-channel structure: the nonstrange isoscalar state
couples strongly to $D^*D^*$, and the strange state has a sizable
coupling to $D^*D_s^*$, whose inclusion is essential for binding. The
doubly strange channel, which belongs exclusively to $\bar{\bm{6}}_S$,
has not yet been calculated, so its entry in Table~V specifies only
the flavor assignment. 

The flavor classification also motivates extending this analysis to
the doubly strange $cc\bar{s}\bar{s}$ system. Such a state, if it
exists, belongs exclusively to $\bar{\bm{6}}_S$. For an $S$-wave
$D_s^*D_s^*$ pair, Bose symmetry allows $J^P=0^+$ and $2^+$. If the
$VV$ channel also provides sufficient attraction in these partial
waves, doubly strange tetraquark states with these quantum numbers may
emerge. Whether such states actually exist must be established through
a consistent coupled-channel dynamical analysis of the
$\bar{\bm{6}}_S$ sector. The corresponding investigation is under
way. 

\section{Summary and conclusions}
\label{sec:4}
We have investigated the $C=+2$, $\mathcal{S}=+1$, and $I=1/2$
two-meson system in an off-shell coupled-channel framework. The
interaction kernel was constructed from $t$- and $u$-channel exchanges
of $\eta$, $K$, $K^*$, $\sigma$, and $f_0(980)$ among the $DD_s$,
$D^*D_s$, $DD_s^*$, and $D^*D_s^*$ channels, without introducing a
bare $s$-channel state. The pseudoscalar- and vector-meson
interactions respect heavy-quark spin symmetry, chiral symmetry, and
hidden local symmetry, while the scalar couplings were determined
from correlated $\pi\pi$ exchange and scalar flavor mixing. We solved
the coupled-channel integral equations in the BbS reduction for
$J^P=0^\pm$, $1^\pm$, and $2^\pm$.

Using a reduced cutoff mass of $\Lambda_0=600$~MeV, we found a
bound-state pole only in the $I(J^P)=\frac12(1^+)$ partial wave, with
its mass $3972.90$~MeV. This corresponds to a binding energy of
$4.01$~MeV relative to the $D^*D_s$ threshold. No other bound-state or
resonance poles were found within the scanned energy region. Since
parity excludes $DD_s$ from the $1^+$ partial wave, the pole lies
below the lowest allowed two-body threshold. It is therefore a bound
state in the present treatment of stable constituent mesons and may be 
interpreted as a molecular $T_{cc\bar s}$.

The comparison with the two-channel subsystems showed that binding
requires the full coupling among $D^*D_s$, $DD_s^*$, and
$D^*D_s^*$. With the same interaction parameters, none of the three
two-channel subsystems generated a pole within the scanned energy
region. Although the $D^*D_s^*$ threshold lies about $148$~MeV above
the bound-state pole, coupling to this channel is essential for
binding. Its sizable ${}^3S_1$ coupling extracted from the pole
residues is consistent with this dynamical role. The substantially
smaller $D$-wave couplings also indicate that the pole couples
predominantly to the $S$-wave components.

The relative signs of the channel couplings determine the dominant
flavor assignment. The ${}^3S_1$ couplings to $D^*D_s$ and $DD_s^*$
were found to have nearly equal magnitudes and opposite signs,
yielding a strong coupling to the antisymmetric combination and a
strongly suppressed coupling to the symmetric one. Together with the
sizable coupling to $D^*D_s^*({}^3S_1)$, these results indicate that
the bound state belongs predominantly to $\bm{3}_A$. This assignment
favors identifying it as the strange partner of the isoscalar
$T_{cc}$ predicted in Ref.~\cite{Kim:2025ado}. Both states exhibit a
significant coupling to the corresponding vector-vector channel. No
additional pole associated with $\bar{\bm{6}}_S$ was found within the
present framework, although this representation also 
contains an $\mathcal{S}=+1$, $I=1/2$ multiplet.

The calculated cross sections exhibited pronounced structures near
the open-charm thresholds. The enhancement in $D^*D_s$ elastic
scattering reflects the nearby bound-state pole, whereas the
structures at the $D^*D_s^*$ threshold arise from threshold cusps
without an additional pole in the scanned region. The nearly
identical line shapes of the $D^*D_s\to D^*D_s^*$ and
$DD_s^*\to D^*D_s^*$ cross sections also reflect the similar dynamics
of the two pseudoscalar-vector channels.

The pole position remains sensitive to the reduced cutoff mass. The
binding energy increased from $4.01$~MeV at $\Lambda_0=600$~MeV to
$15.71$ and $34.04$~MeV at $650$ and $700$~MeV, respectively. As
$\Lambda_0$ decreased, the pole reached the $D^*D_s$ threshold at
$\Lambda_c\simeq557.24$~MeV and continued analytically onto sheet~II
as a virtual-state pole. At $\Lambda_0=550$~MeV, we located it 
$0.264$~MeV below threshold. Thus, the disappearance of the
bound-state pole from the physical sheet does not imply the absence
of a nearby pole. Both its position and its bound-state or
virtual-state character depend on the cutoff, which future
experimental information may help constrain.

The present results indicate that coupled-channel dynamics can
generate a predominantly $\bm{3}_A$ strange partner of the isoscalar
$T_{cc}$. A calculation of its decay properties requires including the
kinematically allowed $DD_s\pi$ and $DD_s\gamma$ channels, which are
omitted here. We are also extending the analysis to the doubly strange
$cc\bar s\bar s$ system. Such a state, if it exists, belongs
exclusively to $\bar{\bm{6}}_S$, and Bose symmetry allows an $S$-wave
$D_s^*D_s^*$ pair with $J^P=0^+$ or $2^+$. Whether the interaction
generates poles in these partial waves must be established through a
consistent coupled-channel calculation. This investigation is under
way.

\begin{acknowledgments}
This work was supported by the Inha University Research Grant (2026).
\end{acknowledgments}

\bibliography{Tccsbar}
\bibliographystyle{apsrev4-1}

\end{document}